\pdfoutput=1
\documentclass[11pt,twoside,a4paper,cmspaper,final,collab]{cms-tdr}
\def\svnVersion{bc16e4f}\def\svnDate{2022/08/12}\def\cmsCernNoTag{CERN-EP-2022-013}\def\cmsCernDate{\today}\def\cmsMessage{Published in the Journal of High Energy Physics as \href{http://dx.doi.org/10.1007/JHEP08(2022)063}{\doi{10.1007/JHEP08(2022)063}.}}
\begin{document}\cmsNoteHeader{SMP-21-002}

\cmsNoteHeader{SMP-21-002}
\title{Measurement of the Drell--Yan forward-backward asymmetry at high dilepton masses in proton-proton collisions at \texorpdfstring{$\sqrt{s} = 13\TeV$}{sqrt(s) = 13 TeV}}

\abstract{
  A measurement of the forward-backward asymmetry of pairs of oppositely charged leptons (dimuons and dielectrons) produced by the Drell--Yan process in proton-proton collisions is presented.  
  The data sample corresponds to an integrated luminosity of 138\fbinv collected with the CMS detector at the LHC at a center-of-mass energy of 13\TeV.  
  The asymmetry is measured as a function of lepton pair mass for masses larger than 170\GeV and compared with standard model predictions. 
  An inclusive measurement across both channels and the full mass range yields an asymmetry of $0.612\pm 0.005\stat\pm 0.007\syst$.
  As a test of lepton flavor universality, the difference between the dimuon and dielectron asymmetries is measured as well. 
  No statistically significant deviations from standard model predictions are observed. 
  The measurements are used to set limits on the presence of additional gauge bosons. 
  For a \PZpr boson in the sequential standard model the observed (expected) 95\% confidence level lower limit on the \PZpr mass is 4.4\TeV (3.7 \TeV). 
}

\newcommand{\MGaMC}{a\textsc{MC@NLO}\xspace}
\newcommand{\muF}{\ensuremath{\mu_\mathrm{F}}\xspace}
\newcommand{\muR}{\ensuremath{\mu_\mathrm{R}}\xspace}
\newcommand{\costheta}{\ensuremath{\cos\theta}\xspace}
\newcommand{\costhetastar}{\ensuremath{\cos\theta^*}\xspace}
\newcommand{\costhetar}{\ensuremath{\cos\theta_\mathrm{R}}\xspace}
\newcommand{\csr}{\ensuremath{c_\mathrm{R}}\xspace}
\newcommand{\Aze}{\ensuremath{A_{0}}\xspace}

\newcommand{\cmsTable}[1]{\resizebox{\textwidth}{!}{#1}}
\newlength\cmsTabSkip\setlength{\cmsTabSkip}{1ex}

\hypersetup{
pdfauthor={CMS Collaboration},
pdftitle={Measurement of the Drell-Yan forward-backward asymmetry at high dilepton masses in proton-proton collisions at sqrt(s) = 13 TeV},
pdfsubject={CMS},
pdfkeywords={CMS, Drell-Yan, AFB, asymmetry, Z prime }}

\maketitle

\section{Introduction} \label{sec:intro}

Drell--Yan (DY) production of pairs of same-flavor, oppositely charged leptons ($\Pell^+\Pell^-$) in proton-proton ($\Pp\Pp$) collisions 
occurs via the $s$-channel exchange of $\PZ/\PGg^*$ bosons.
At leading order (LO) in quantum chromodynamics (QCD), the $\PZ/\PGg^*$ bosons are produced via quark-antiquark (\qqbar) annihilation.
The presence of both vector and axial couplings of gauge bosons to leptons results in a forward-backward asymmetry, \AFB,
in the angular distribution of the final-state lepton with respect to the initial-state quark. 
Deviations from the standard model (SM) predictions of \AFB can result from the presence of an additional 
neutral gauge boson (\PZpr)~\cite{London:1986, Rosner:1987a, Rosner:1996, Bodek:2001xk, Rizzo:2006nw, Accomando:2015cfa, Accomando:2019ahs},
quark-lepton compositeness~\cite{Eichten:1983hw}, large extra dimensions~\cite{Hewett:1998sn}, vector-like fermions~\cite{Gross:2016ioi}, 
certain dark matter candidates~\cite{Capdevilla:2017doz}, or leptoquarks~\cite{Raj:2016aky}.
Hints of a violation of lepton flavor universality in several measurements recently reported by the LHCb Collaboration~\cite{LHCb:2017avl, LHCb:2019hip, LHCb:2021trn}
have sparked interest in models for physics beyond the SM that could explain these effects. 
Many of these models include heavy neutral gauge bosons or leptoquarks with non-flavor-universal couplings~\cite{Allanach:2020kss, Becirevic:2016oho}, 
which could produce signatures in high-mass dilepton distributions~\cite{Greljo:2017vvb}.
As compared to measurements of differential cross sections at high dilepton masses, 
measurements of \AFB have a reduced dependence on systematic uncertainties related to the reconstruction and identification of high-momentum leptons. 
Hence, precision measurements of \AFB can provide stringent tests of the electroweak sector of the SM and of lepton flavor universality.
The main results of this paper are measurements of \AFB as a function of dilepton mass for both muons and electrons in the high-mass region ($>$170\GeV),
which is particularly sensitive to potential contributions from new physics.

The asymmetry is defined in terms of the angle $\theta$ between the negatively charged final-state lepton and the initial-state quark (meant here in contrast to the antiquark)
in the $\Pell^+\Pell^-$ center-of-mass frame,
\begin{equation}
  \label{eq:afb_def}
  A_{\mathrm{FB}} = \frac{\sigma_\mathrm{F} - \sigma_\mathrm{B}}{\sigma_\mathrm{F} + \sigma_\mathrm{B}},
\end{equation}
where $\sigma_\mathrm{F}$ ($\sigma_\mathrm{B}$) is the total cross section for forward (backward) events, defined by $\costheta > 0$ ($<$0). 
The sign of $\theta$ is defined so that $\costheta = 1$ events are those in which the negatively charged final-state lepton is traveling 
in the same direction as the incident quark.
The \AFB value can also be directly related to the parton-level differential cross section.
The latter can be written as~\cite{Mirkes:1992hu,Mirkes:1994dp}:
\begin{equation}
  \label{eq:ang_xsec}
  \dd{\sigma}{\cos\theta}\propto \frac{3}{8}\biggl[ 1 + \cos^2\theta + \frac{\Aze}{2} \left(1-3 \cos^2\theta\right) + A_{4}\costheta\biggr],
\end{equation}
where \Aze and $A_4$ are the standard dimensionless constants parameterizing the angular distribution of the DY process~\cite{Mirkes:1992hu,Mirkes:1994dp}. 
The terms that are functions of \costheta with even parity do not contribute to \AFB, leading to the relation:
\begin{equation}
  \label{eq:a4_afb}
  \frac{3}{8} A_4 = A_{\mathrm{FB}}.
\end{equation}
This form of the angular distribution is general for any $s$-channel processes mediated by a \mbox{spin-1} boson and 
is thus applicable in the case of interference from additional heavy vector bosons.
The angular coefficients \Aze and $A_4$ vary as functions of the mass ($m$), transverse momentum (\pt), and rapidity ($y$) of the dilepton system. 
As the dilepton transverse momentum approaches zero, \Aze vanishes.
It is nonzero for finite  $\PZ/\PGg^*$ \pt caused by processes in which $\PZ/\PGg^*$ bosons are produced in association with additional jets. 
Thus, measurements of \Aze probe higher-order corrections in perturbative QCD.
Additional results of this paper are the measurements of \Aze as a function of the dilepton mass in the same high-mass region ($>$170\GeV).

Electroweak interference between the photon and the \PZ boson leads to
negative values of \AFB with large absolute value for masses below the \PZ pole ($m < 80\GeV$), and leads to large and positive values of \AFB above the \PZ pole ($m > 110\GeV$). 
Near the \PZ boson mass peak, \AFB reflects pure \PZ exchange and is close to zero because of the small value of the charged-lepton vector coupling to \PZ bosons.
In the high-mass region, above the \PZ boson peak, the SM value of \AFB is approximately constant with a value of $\approx$0.6. 
The \Aze coefficient is expected to be $\approx$0.06 at 170 \GeV and should decrease at higher masses.

For the case of a new heavy \PZpr, off-shell interference can produce deviations at masses significantly lower than the \PZpr mass. 
The deviation from the SM \AFB is insensitive to the width of the \PZpr~\cite{Accomando:2015cfa, Accomando:2019ahs}.
This measurement thus offers a complementary approach to previous searches for new physics in the dilepton channel that have searched for a peak in the invariant mass distribution 
caused by the resonant production of a new particle~\cite{Aad:2019fac, CMS:2021ctt}. 
Additionally, because the constraining power of this technique is based on interference rather than direct production, 
its sensitivity to higher mass scales is not limited by the center-of-mass energy and will continuously improve with the increased statistical precision with the addition of future LHC data. 

When the dilepton system has nonzero \pt, the exact directions of the incident partons are unknown since they are no longer collinear with the proton beams.
To minimize the impact of this effect on the asymmetry measurement, the Collins--Soper rest frame~\cite{Collins:1977} is used. 
In this frame, $\theta^*$ is defined as the angle between the negatively charged lepton and the axis that bisects the angle between the incident parton directions. 
Approximating the leptons as massless, \costhetastar can be computed from lab frame variables as:
\begin{equation}
  \label{eq:cs}
  \costhetastar = \frac{2(P_{1}^{+} P_{2}^{-} - P_{1}^{-}P_{2}^{+})}{ m \sqrt{m^2 + \pt^2}},
\end{equation}
with
\begin{equation*}
  P^{\pm}_{i} = \frac{1}{\sqrt{2}}(E_i \pm p_{z,i}),
\end{equation*}

where $E_i$ is the energy and $p_{z,i}$ the longitudinal momentum of the lepton ($i=1$) and antilepton ($i=2$).
The sign of \costhetastar is defined with respect to the direction of the incident quark,
which is unknown for collisions in a $\Pp\Pp$ collider.
Typically, the quarks in the collision will carry a larger momentum fraction than the antiquarks, since only the quarks are valence partons of protons.
Thus, usually the lepton pair will have longitudinal momentum along the direction of the incident quark. 
Therefore, instead of using the initial quark direction, one can define the positive axis to be the longitudinal direction of the lepton pair. 
We denote the angular variable defined in this way as: 
\begin{equation}
  \costhetar  = \frac{p_z}{\abs{p_z}} \costhetastar.
\end{equation}

where $p_z$ is the longitudinal momentum of the dilepton system.
The presence of events where the quark direction does not match the lepton pair direction dilutes the asymmetry observed in \costhetar 
as compared with the underlying asymmetry in \costhetastar.
The fraction of events for which the quark direction is the same as the longitudinal momentum of the lepton pair increases with the absolute value of the dilepton rapidity.
At an invariant mass of 500\GeV, this fraction is $\approx$60\% at $\abs{y} = 0.2$ and $\approx$95\% at $\abs{y} = 2$.

The asymmetry, \AFB, was previously measured by the CMS Collaboration at $\sqrt{s} = 7\TeV$~\cite{CMS-PAPERS-EWK-11-004} and 8\TeV~\cite{CMS-PAPERS-SMP-14-004} 
and by the ATLAS Collaboration at $\sqrt{s} = 7\TeV$~\cite{Aad:2015uau}.
Around the \PZ boson peak, measurements of \AFB have been used by CMS and ATLAS to extract 
the effective weak mixing angle $\mathrm{sin}^{2}\theta^{\ell}_\mathrm{eff}$~\cite{CMS:2018ktx,Aad:2015uau}.
The results presented in this paper at $\sqrt{s} = 13\TeV$ takes a new approach to measuring \AFB. 
Rather than counting forward and backward events, \AFB is extracted by fitting a set of templates 
constructed to represent the different terms in Eq.~(\ref{eq:ang_xsec}) to the measured angular distribution. 
These templates are constructed from Monte Carlo (MC) simulations, which include the dilution effect.
The \AFB measured by the template fitting method corrects for the dilution effect and thus determines the asymmetry in \costhetastar,
whereas the \AFB measured by the traditional counting method does not correct for the dilution and determines the asymmetry in \costhetar. 
Additionally, the template-based measurement of \AFB optimally combines the information from the full fiducial dilepton rapidity range into a single measurement, 
rather than being reported differentially in rapidity as is done in the counting method.
This means that the \AFB values from this measurement are not directly comparable with the previous CMS results.
The previous ATLAS publication utilized the counting method but also included additional results unfolding the dilution effect,
that are comparable with this measurement.
This template-based measurement is a maximum-likelihood estimate of \AFB and leads to a $\approx$20\% smaller uncertainty than a simple counting-based measurement \cite{Bodek:2010qg}. 
However it relies more heavily on parton distribution functions, and cannot be as easily reinterpreted using new models as a counting-based measurement of the diluted \AFB.

\section{Analysis strategy}\label{sec:analysis_strategy}

The observed distribution of the reconstructed scattering angle, \costhetar, abbreviated as \csr, can be expressed as a convolution of the \costhetastar, abbreviated $c_{*}$, 
distribution defined in Eq.~(\ref{eq:ang_xsec}),
\begin{equation}
  f(\csr) = C \int \rd c_* R(\csr; c_* )
  \varepsilon (c_*)\dd{\sigma}{c_*},
\end{equation}
where $C$ is a normalization constant, $R$ is a ``reconstruction function'' that incorporates detector resolution and the dilution effect, 
and $\varepsilon$ is an efficiency function.  
It is important to point out that convolutions are linear operations, which implies that the resulting reconstructed distribution can be expressed as a sum of convolutions of each of the terms
in Eq.~(\ref{eq:ang_xsec}).

This linearity allows the fitting function to be represented by a set of parameter-independent templates corresponding to the different terms in Eq.~(\ref{eq:ang_xsec}).  
By fitting the observed data distribution to combinations of these signal templates and additional background templates, one can extract \AFB and \Aze.
The DY signal templates, representing the reconstructed angular distribution corresponding to each of the terms in the true angular distribution of Eq.~(\ref{eq:ang_xsec}), 
are constructed by reweighting MC simulated events.
The details of the signal template construction are discussed in Section~\ref{sec:temp_construct}.

The templates for the dominant backgrounds are extracted from MC simulation and validated in a control region of $\Pe\Pgm$ events.
Templates for additional backgrounds that are not well modeled by MC are constructed based on control samples in data.

Because the form of Eq.~(\ref{eq:ang_xsec}) is general for any $s$-channel spin-1 process, the measured values of \AFB can be used 
to set limits on the existence of heavy \PZpr bosons.

To search for signs of lepton flavor universality violation, the difference between \AFB and \Aze in the muon and electron channels can be measured directly.
The parameters of interest in these measurements are $\Delta\AFB = A_{\mathrm{FB},\PGm\PGm} - A_{\mathrm{FB,ee}}$ and 
$\Delta\Aze = \mathrm{A}_{0,\PGm\PGm} - \mathrm{A}_{0,\Pe\Pe}$.
The direct measurement of $\Delta\AFB$ is less sensitive to systematic uncertainties common to the muon and electron channels than the measurement of \AFB in the individual channels,
and thus has reduced systematic uncertainty.

\subsection*{Fiducial corrections and measurement interpretation}

In this analysis, \AFB and \Aze are measured as functions of the dilepton mass, but \AFB and \Aze also depend on the dilepton \pt and rapidity.
A single \Aze parameter and a single \AFB parameter are measured in each mass bin, representing effective values, integrated over \pt and rapidity.
However, because the acceptance of dilepton events in the CMS detector is not independent of the dilepton $\pt$ and rapidity, the effective \Aze or \AFB
of recorded events is not the same as the effective value in the full phase space. 
The effect was studied in MC simulations and it was found that differences between the fiducial and full phase space values of \Aze were insignificant,
but the differences in \AFB could be as large as 2\% in the lower mass bins. 
Using these MC simulations, a correction factor was derived to convert the fiducial \AFB directly measured in the data to that of the full phase space. 
This correction factor is then applied to the result of the template fit so that the final reported value is a measurement of the \AFB in the full phase space.
Uncertainties in the MC simulation are propagated through the evaluation of this correction factor, 
and are treated as a source of systematic uncertainty in the final measurement

This template fitting technique automatically accounts for all other resolution, dilution, migration, and acceptance effects since they are modeled in the simulation.
Uncertainties in the simulation of these effects can be accounted for using variations of the corresponding templates, as discussed in Section~\ref{sec:sys_uncs}.
The final result can thus be interpreted as a measurement of the partonic \AFB in different bins of the reconstructed lepton pair mass.
Because the mass bins used are large and the SM \AFB is essentially independent of the dilepton mass in the high-mass region of our measurement,
the measurement is not unfolded to parton-level mass bins.

\section{The CMS detector and physics objects} \label{sec:detector}
The central feature of the CMS apparatus is a superconducting solenoid of 6\unit{m} internal diameter, providing a magnetic field of 3.8\unit{T}. 
Within the solenoid volume are a silicon pixel and strip tracker, a lead tungstate crystal electromagnetic calorimeter (ECAL), and a brass and scintillator hadron calorimeter (HCAL), 
each composed of a barrel and two endcap sections. 
Forward calorimeters extend the pseudorapidity ($\eta$) coverage provided by the barrel and endcap detectors. 
Muons are detected in gas-ionization chambers embedded in the steel flux-return yoke outside the solenoid.
A more detailed description of the CMS detector, together with a definition of the coordinate system used and the relevant kinematic variables, is reported in Ref.~\cite{Chatrchyan:2008zzk}.

Events of interest are selected using a two-tiered trigger system. 
The first level, composed of custom hardware processors, uses information from the calorimeters and muon detectors 
to select events at a rate of around 100\unit{kHz} within a fixed latency of about 4\mus~\cite{Sirunyan:2020zal}. 
The second level, known as the high-level trigger, consists of a farm of processors running a version of the full event reconstruction software 
optimized for fast processing, and reduces the event rate to around 1\unit{kHz} before data storage~\cite{Khachatryan:2016bia}.

The particle-flow algorithm~\cite{Sirunyan:2017ulk} aims to reconstruct and identify each individual particle in an event, 
with an optimized combination of information from the various elements of the CMS detector. 
The energy of photons is obtained from the ECAL measurement. 
The energy of electrons is determined from a combination of the electron momentum at the primary interaction vertex as determined by the tracker, 
the energy of the corresponding ECAL cluster, and the energy sum of all bremsstrahlung photons spatially compatible with originating from the electron track. 
The energy of muons is obtained from the curvature of the corresponding track. 
The energy of charged hadrons is determined from a combination of their momentum measured in the tracker and the matching ECAL and HCAL energy deposits, 
corrected for the response function of the calorimeters to hadronic showers. 
Finally, the energy of neutral hadrons is obtained from the corresponding corrected ECAL and HCAL energies.
The candidate vertex with the largest value of summed physics-object $\pt^2$ is assigned to be the primary $\Pp\Pp$ interaction vertex. 
The physics objects are the jets, clustered using the anti-\kt jet finding algorithm~\cite{Cacciari:2008gp,Cacciari:2011ma} with the tracks assigned to candidate vertices as inputs.

Muons are measured in the range $\abs{\eta} < 2.4$, with detection planes made using three technologies: drift tubes, cathode strip chambers, and resistive-plate chambers. 
The single-muon trigger efficiency exceeds 90\% over the full $\eta$ range, and the efficiency to reconstruct and identify muons is greater than 96\%. 
Matching muons to tracks measured in the silicon tracker results in a relative \pt resolution of 1\% in the barrel and 3\% in the endcaps for muons with \pt up to 100\GeV,
and of better than 7\% for muons in the barrel with \pt up to 1\TeV~\cite{Sirunyan:2018}.

The single-electron trigger efficiency is approximately 80\% over the full $\eta$ range, 
and the efficiency to reconstruct and identify electrons is greater than 65\% for electrons with $\pt > 20\GeV$. 
The momentum resolution for electrons with $\pt \approx 45\GeV$ from $\Z \to \Pe\Pe$ decays ranges from 1.7--4.5\%. 
It is generally better in the barrel region than in the endcaps, and also depends on the bremsstrahlung energy emitted by the electron as it traverses the material in front of the ECAL~\cite{CMS:2020uim}.

Jets are clustered from the particle-flow candidates in an event using the anti-\kt jet finding algorithm with a distance parameter of 0.4.
Jet momentum is determined as the vectorial sum of all particle momenta in the jet, and is found from simulation to be, on average, within 5--10\% 
of the true momentum over the whole \pt spectrum and detector acceptance. 
Additional $\Pp\Pp$ interactions within the same or nearby bunch crossings ("pileup") 
can contribute additional tracks and calorimetric energy depositions, increasing the apparent jet momentum. 
To mitigate this effect, tracks identified as originating from pileup vertices are discarded and an offset correction is applied to correct for remaining contributions~\cite{Sirunyan_2020pileup}. 
Jet energy corrections are derived from simulation studies so that the average measured energy of jets becomes identical to that of particle level jets~\cite{Khachatryan:2016kdb}. 
In situ measurements of the momentum balance in dijet, photon+jet, Z+jet, and multijet events are used to determine any residual differences between the jet energy scale in data and in simulation,
and appropriate corrections are made~\cite{Khachatryan:2016kdb}. 
The missing transverse momentum vector (\ptvecmiss) is defined as the negative vector \pt sum of all the particle-flow candidates in an event, 
and its magnitude is denoted as \ptmiss~\cite{Sirunyan:2019kia}. 
The \ptvecmiss is modified to account for corrections to the energy scale of the reconstructed jets in the event.

\section{Data and Monte Carlo samples} \label{sec:data_mc_samples}
The analysis is performed with $\Pp\Pp$ collision data collected with the CMS detector at the LHC in 2016--2018 at $\sqrt{s} = 13\TeV$ . 
The total integrated luminosity amounts to 138\fbinv~\cite{CMS:2021xjt, CMS:2018elu, CMS:2019jhq}.

{\tolerance=800 Various MC generators have been used to simulate the DY signal and background processes. 
The DY signal samples, $Z/\PGg* \to \Pe\Pe$ and $Z/\PGg^{*} \to \Pgm\Pgm$, and the background sample $Z/\PGg^{*} \to \PGt\PGt$,
have been generated at next-to-leading order (NLO) with \MGvATNLO~\cite{Alwall:2014hca} (shortened as \MGaMC),
using version v2.2.2 (v2.6.0) for samples corresponding to the 2016 (2017--2018) data-taking period.
Up to two additional partons are allowed using the FxFx merging scheme~\cite{Frederix:2012ps}. 
The samples are interfaced with \PYTHIA~\cite{Sjostrand:2014zea} to simulate the parton shower, hadronization, and quantum electrodynamics final-state radiation.
The CUETP8M1~\cite{CMS:2015wcf} (CP5~\cite{CMS:2019csb}) \PYTHIA tune and \PYTHIA version v8.226 (v8.230) are used for the samples corresponding to the 2016 (2017--2018) data-taking period.
The NLO NNPDF 3.0 parton distribution functions (PDFs)~\cite{Ball:2010de, Ball:2014uwa} are used for all three data-taking periods.\par}

Other processes that can give a final state with two oppositely charged same-flavor leptons are diboson production 
($\PW\PW$, $\PW\PZ$, $\PZ\PZ$), photon-induced dilepton production ($\PGg \PGg \to \Pell \Pell$), 
top quark pair production (\ttbar), and single top quark production in association with a $\PW$ boson ($\PQt\PW$).
The \ttbar and $\PQt\PW$ backgrounds are generated at NLO using \POWHEG v2.0~\cite{Nason:2004rx,Frixione:2007vw,Alioli:2010xd,Frixione:2007nw}, 
and interfaced to \PYTHIA with 
the CUETP8M2T4~\cite{CMS-PAS-TOP-16-021} (CP5) tune for the 2016 (2017--2018) data-taking period.
Background samples of $\PZ\PZ \to \Pell \Pell \Pell' \Pell'$, $\PZ\PZ \to \Pell \Pell \PGn \PGn$, and $\PW\PW \to \Pell \PGn \Pell \PGn$
are generated at NLO with \POWHEG interfaced to \PYTHIA.
Background samples of $\PW\PZ \to \PQq\PQq \Pell \Pell$ and $\PZ\PZ \to \PQq\PQq \Pell \Pell $ are generated at NLO with \MGaMC
interfaced to \PYTHIA.
The $\PW\PW$, $\PW\PZ$, and $\PZ\PZ$ samples corresponding to the 2016 (2017--2018) data-taking period are interfaced with \PYTHIA using the CUETP8M1 (CP5) tune. 
The $\ttbar$, $\PQt\PW$, and diboson backgrounds corresponding to the 2016 (2017--2018) data-taking period use the NNPDF 3.0 (3.1) PDFs.
The photon-induced background,
$\PGg \PGg \to \Pell \Pell$, is simulated using the \textsc{CepGen}~\cite{Forthomme:2018ecc} implementation of \textsc{LPAIR}~\cite{Vermaseren:1982cz,Baranov:1991yq},
interfaced to \PYTHIA v6.429~\cite{Sjostrand:2006za}, and using the default proton structure function parameterization of Suri--Yennie~\cite{Suri:1971yx}.
This contribution is split into three parts because the interaction at each proton vertex
can be elastic or inelastic.

The cross sections of $\PW\PZ$ and $\PZ\PZ$ diboson samples are normalized to the NLO predictions calculated
with \textsc{MCFM~6.6}~\cite{Campbell:2015qma}, whereas the cross sections of the $\PW\PW$
samples are normalized to the next-to-NLO (NNLO) predictions~\cite{Gehrmann:2014fva}. The total cross section of the \ttbar
process is normalized to the prediction with NNLO accuracy in QCD and next-to-next-to-leading-logarithmic accuracy for the soft gluon radiation
resummation calculated with \TOPpp2.0~\cite{Czakon:2011xx}.

The detector response for all MC samples is simulated using a detailed description of the CMS detector based on \GEANTfour~\cite{AGOSTINELLI2003250}. The pileup distribution in simulation is weighted to match the one observed in data.

\section{Event selection} \label{sec:event_selection}
Events are required to have two leptons of the same flavor and opposite charges.
The dimuon and dielectron events are selected by single-muon and single-electron triggers, respectively.
The thresholds of these triggers are different for the different data-taking years, 
and the leading muon or electron in the event is required to have a \pt above the trigger threshold.
In the analysis, the leading muon \pt requirement for the years 2016/2017/2018 is 26/29/26\GeV and for electrons it is 29/38/35\GeV.
The subleading lepton is required to have $\pt > 15\GeV$.
All muons are required to be within the acceptance of the muon system ($\abs{\eta} < 2.4$) and all electrons must be within $\abs{\eta} < 2.5$, excluding the barrel-endcap transition
region of the ECAL ($1.44 < \abs{\eta} < 1.57$).
Additionally, to remove cosmic-ray--induced events, the azimuthal angle ($\phi$) between the two muons is required to differ from $\pi$ by more than 5\unit{mrad}.

Each reconstructed muon is required to pass identification criteria that are based on the number of hits observed in the tracker, the response of the muon detectors,
and a set of matching criteria between muon track parameters, as measured by the inner tracker and muon detectors. 
To suppress nonprompt muons coming from heavy-flavor decays, both muons must be isolated from other particles in a cone of size $\Delta R = 0.3$,
where $\Delta R = \sqrt{\smash[b](\Delta \eta)^2 + (\Delta \phi)^2}$ refers to the distance from the muon to a given track. 
More details on the muon identification and reconstruction used in this analysis can be found in Refs.~\cite{Sirunyan:2018,CMS:2019ied}. 

The reconstructed electron candidates are required to pass identification criteria that are based on electromagnetic shower shape variables. 
Electrons originating from photon conversions are suppressed by requiring that the candidates have at most one missing inner tracker hit 
and are not consistent with being part of a conversion pair. 
Electrons are also required to be isolated from other particles within a cone of size $\Delta R = 0.3$. 
The electron isolation criteria are based on the ratio of the electron \pt to the sum of energy deposits associated with the
photons as well as with the charged and neutral hadrons reconstructed by the particle-flow algorithm.
More details on the electron reconstruction and identification criteria used in this analysis
are described in Ref.~\cite{CMS:2020uim}.

To suppress backgrounds that contain the decays of top quarks, two additional requirements are applied.
First, events are required to have $\ptmiss < 100\GeV$.
Second, it is required that neither of the two highest \pt jets in the event with $\abs{\eta} < 2.4$ are identified as a bottom quark jet (\PQb tagged).
Jets originating from decays of bottom quarks are identified using an algorithm that combines lifetime information
from tracks and secondary vertices~\cite{BTV-16-002}. 
A working point is used that has a 68\% efficiency of correctly identifying a bottom quark jet and a 1\% probability of misidentifying a light-flavor quark
or gluon jet as a bottom quark jet.

\section{Backgrounds} \label{sec:backgrounds}

At large lepton pair masses, the dominant background comes from fully leptonic decays of \ttbar events. 
There are also backgrounds from $\PQt\PW$ events, diboson processes and 
$\PGt \PGt$ leptonic decays.
All of these backgrounds are well modeled in simulation, and the estimated event yields are validated in data in a control region of $\Pe\Pgm$ events.
Multijet and $\PW$+jets events where one or more jets are incorrectly identified as leptons (denoted as ``MisID'' events) 
are also a source of background.
This background is larger for the $\Pe\Pe$ channel than for the $\Pgm\Pgm$ channel. 
A technique based on control samples in data is employed to estimate this background. 
There are also backgrounds coming from $t$-channel photon-induced dilepton production, which are modeled with MC simulation as well.

The MisID background is estimated from data using the ``misidentification rate'' method.
Descriptions of this method are reported in Refs.~\cite{CMS:2018mdl,CMS:2014lcz}.
The misidentification rate is defined as the probability of a jet, having been reconstructed as a lepton candidate,
to pass the lepton selection requirements.
This rate is measured in a sample with
two leptons coming from a \PZ boson decay and an additional, potentially misidentified, lepton.  
These two leptons are required to pass the lepton identification requirements and 
have an invariant mass within 7\GeV of the \PZ boson mass~\cite{ParticleDataGroup:2020ssz}.
The presence of the third lepton candidate is used as a probe to measure the misidentification rate.
These lepton candidates are required to pass a set of identification and isolation requirements
less stringent than the full selection requirements of the analysis.
The MisID background can then be estimated from a sample of data events with two lepton 
candidates where at least one of the candidates fails the full selection requirements. 
Events from this sample are assigned weights based on the expected misidentification probability of
the failing lepton candidates. 
Contamination of this sample by lepton pairs from DY and other prompt processes is subtracted using MC simulation.
These reweighted events are used to estimate the yield and shape of the MisID background. 

This method of background estimation is validated in a control region.
This control region has all the same selection criteria and covers the same dilepton mass range ($m > 170\GeV$) as the signal region 
except the lepton pairs are required to have same-sign rather than opposite-sign charges.
A large fraction of the events in this control region stem from misidentified jets.
In the $\Pe\Pe$ channel, there is also a significant contribution to this sample from opposite-sign DY events where one of the electrons has 
had its charge incorrectly assigned. 
The rate of this misassignment is not modeled well in simulation, and so a correction to the MC is derived in the mass window of the \PZ peak ($70 < m < 110\GeV$)
and applied.
The MisID background estimate as well as MC estimates of other backgrounds are compared with the observed yield of same-sign events.
A \costhetar-dependent correction to the MisID background estimate is derived using the ratio of the MisID estimate to the number of observed same-sign events minus other backgrounds.
The uncertainty in this correction is calculated as the quadratic sum of the statistical uncertainties from the limited sample size,
uncertainty in the amount of DY pairs reconstructed as same-sign pairs, 
as well as systematic uncertainty reflecting possible differences in shape between same-sign and opposite-sign MisID estimates.

\begin{figure}[hbtp]
  \centering
  \includegraphics[width = 0.49 \textwidth]{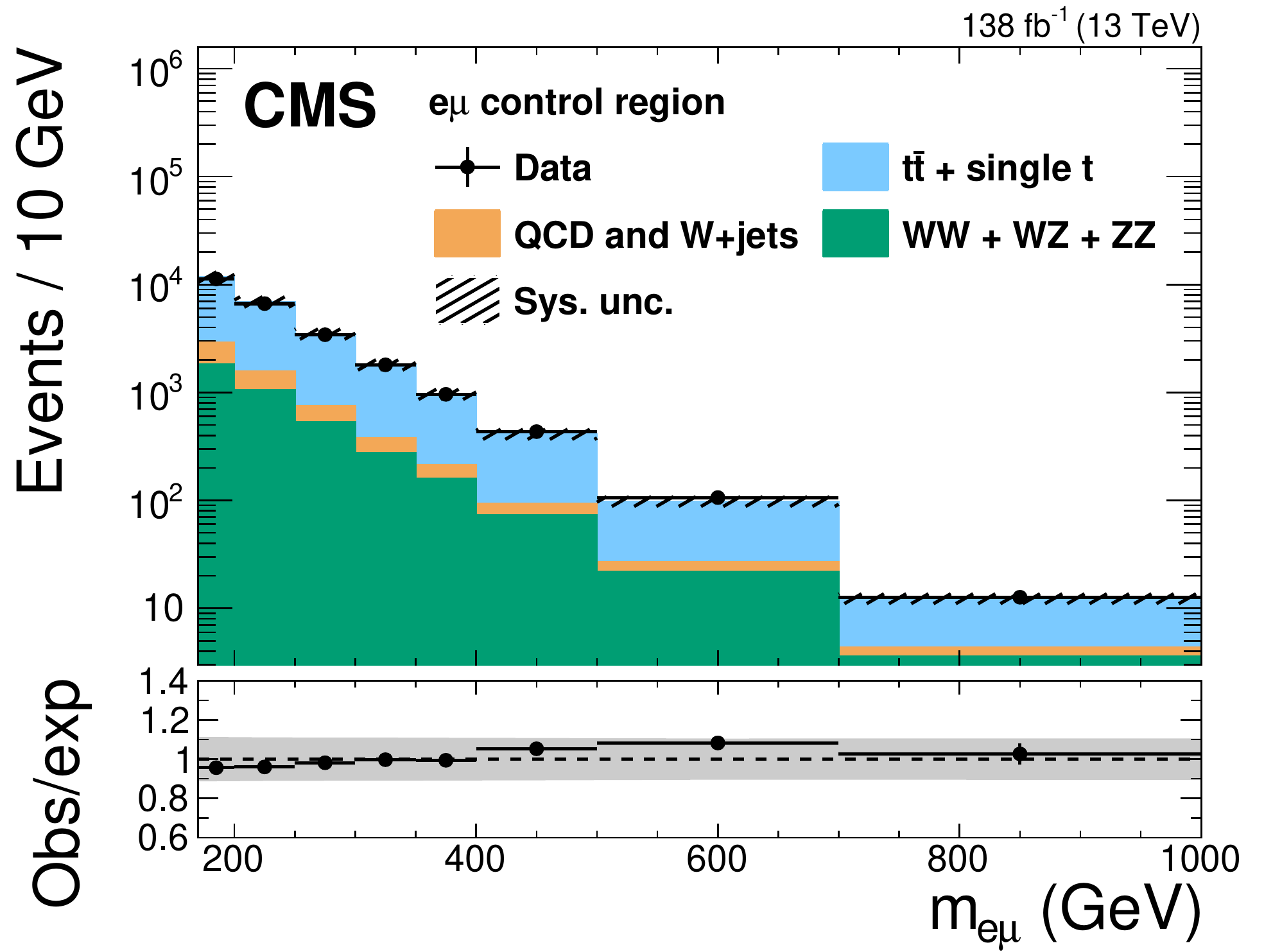}
  \includegraphics[width = 0.49 \textwidth]{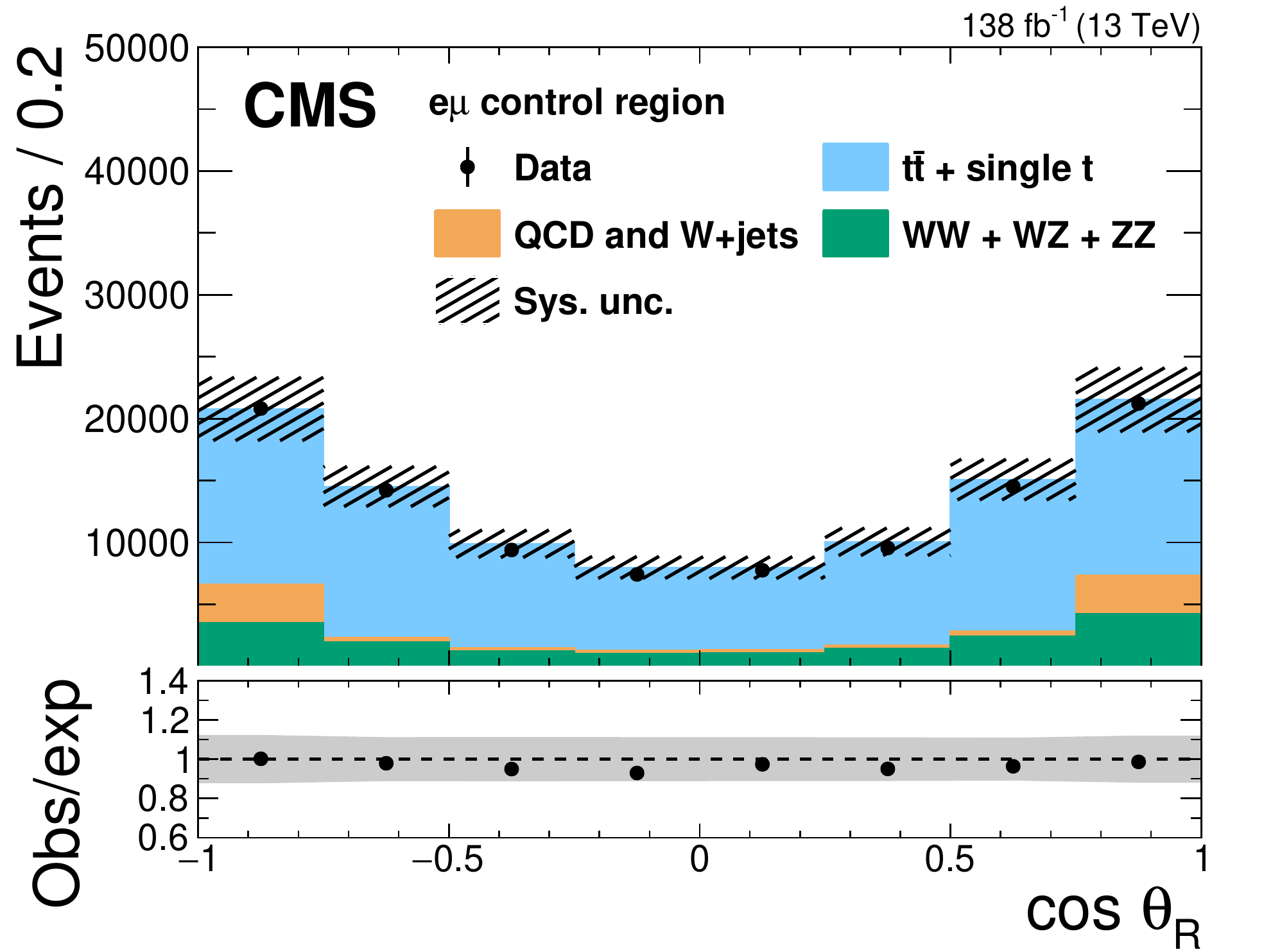}
  \caption{The invariant mass distribution (left) and \costhetar distribution (right) of $\Pe\Pgm$ events observed in CMS data (black dots with statistical uncertainties)
    and expected backgrounds (stacked histograms).
    The hatched bands show the systematic uncertainty in the expected yield. The sources of this uncertainty are discussed in Section~\ref{sec:sys_uncs}.
    The lower panels show the ratio of the data to the expectations. The gray bands represent the total uncertainty in the predicted yields.}
  \label{fig:emu}
\end{figure}

Because the $\PGt \PGt$, $\PQt\PW$, \ttbar, and diboson backgrounds all have decays to $\Pe\Pgm$ pairs as well, the MC simulation
of these processes can be validated in data.
The $\Pe\Pgm$ control region used has all the same selection criteria as the signal region, except that events are required to have one muon and one electron
rather than a pair of the same flavor.
The muon is required to have ${\pt > 26/29/26\GeV}$ for the years 2016/2017/2018 and pass the muon trigger used in the main analysis, 
and the electron is required to have ${\pt > 15\GeV}$. 
There are also MisID events in this $\Pe\Pgm$ control region coming from QCD multijet and W+jets backgrounds. 
These are estimated using the misidentification rate technique previously described. 
The dilepton invariant mass and \costhetar distributions of $\Pe\Pgm$ events are shown in Fig.~\ref{fig:emu}.
Good agreement is observed between the simulated and observed yield of $\Pe\Pgm$ events across the entire mass range.

Because diboson events are produced via electroweak processes, they are expected to have a small forward-backward asymmetry in their lepton 
pairs; \ttbar events are also known to have an asymmetry, but it is too small to be detected in its lepton pairs alone~\cite{Sirunyan:2019eyu}.
The QCD and W+jets backgrounds should have no asymmetry. 
Based on MC and control sample estimates, we predict an overall forward-backward asymmetry of $\approx$0.01 in
the sample of $\Pe\Pgm$ events. 
Using the definition in Eq.~(\ref{eq:afb_def}), the \AFB of the $\Pe\Pgm$ sample is found to be $0.012\pm0.003$, consistent with this expectation.

Although the overall normalizations and asymmetries of the MC $\Pe\Pgm$ estimates are observed to be consistent with data, 
there are discrepancies between the predicted and observed shapes of the \costhetar distribution in certain ranges of dilepton mass. 
To address these discrepancies, a \costhetar correction is derived from the $\Pe\Pgm$ sample in data, 
based on the ratio of the observed and predicted \costhetar distributions. 
Because the asymmetry is modeled well, this correction is derived symmetrically in \costhetar using four bins of $\abs{\costhetar}$.
To account for the changing \costhetar shapes as a function of the dilepton mass, the correction is derived separately in different mass bins. 
Corrections are derived in five mass bins (170, 200, 250, 320, 510, 3000\GeV) with edges matching those used in the \AFB measurement (defined in Section~\ref{sec:temp_construct}),
combining the highest mass bins because of a limited event count.
These corrections are applied to modify the shapes of the templates used to model the corresponding backgrounds in the signal region.
They change the estimated \costhetar shape of these backgrounds and introduce uncertainties in these shapes at the level of by $\approx5\pm4\%$ and $\approx15\pm10\%$  
in the lowest and highest mass bins, respectively.

\begin{figure}[hbtp]
  \centering
  \includegraphics[width = 0.49 \textwidth]{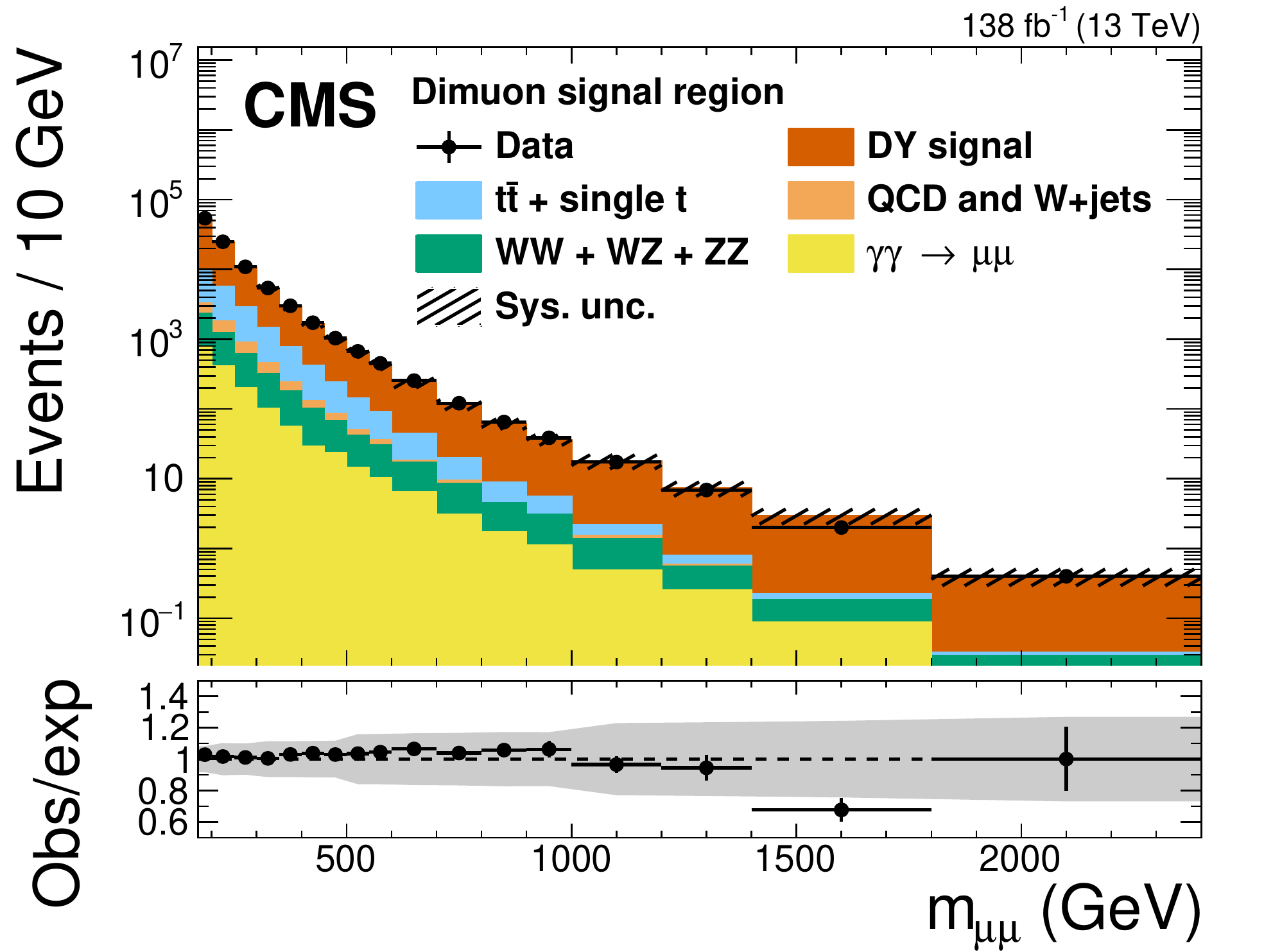}
  \includegraphics[width = 0.49 \textwidth]{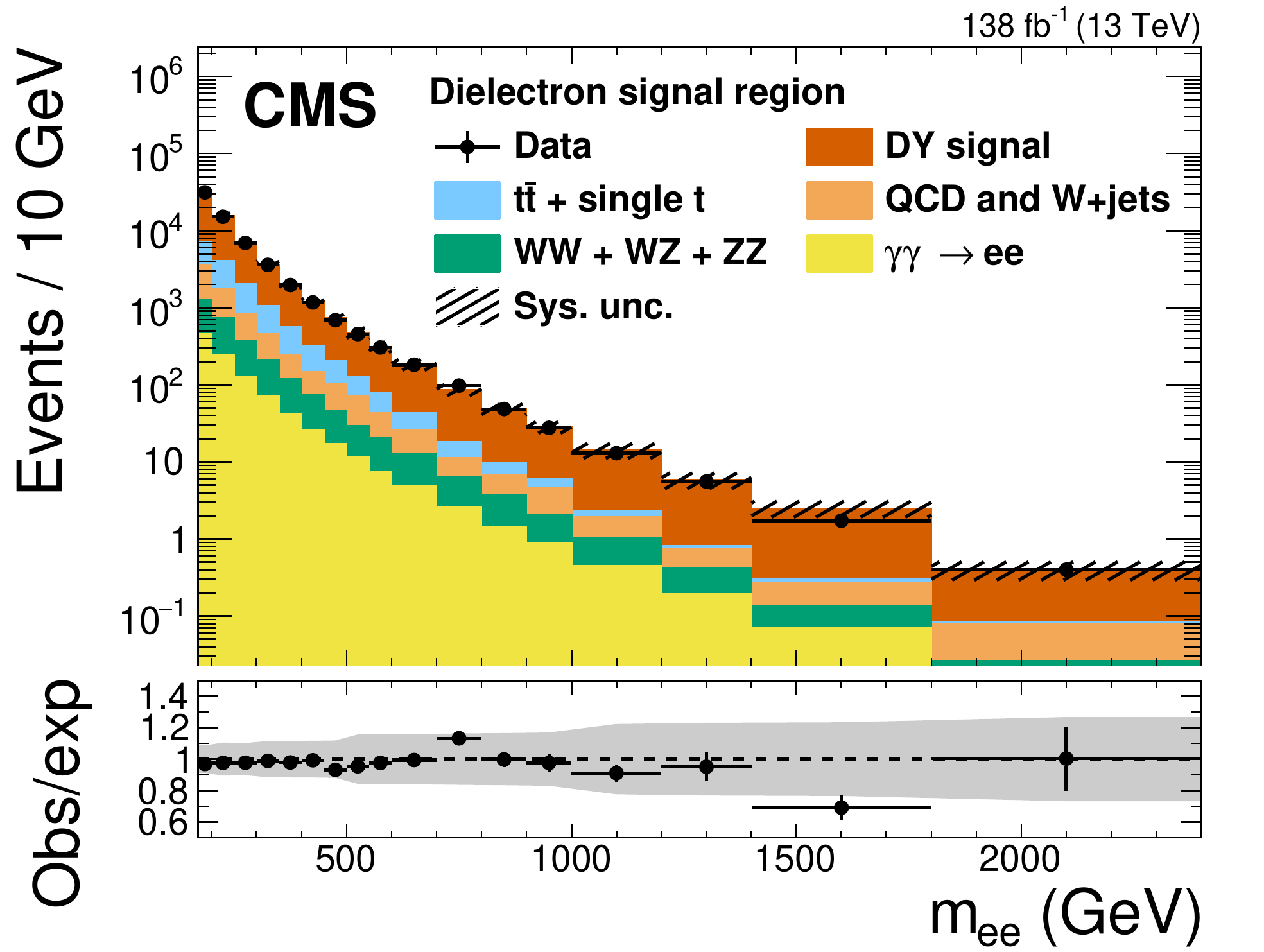}
  \includegraphics[width = 0.49 \textwidth]{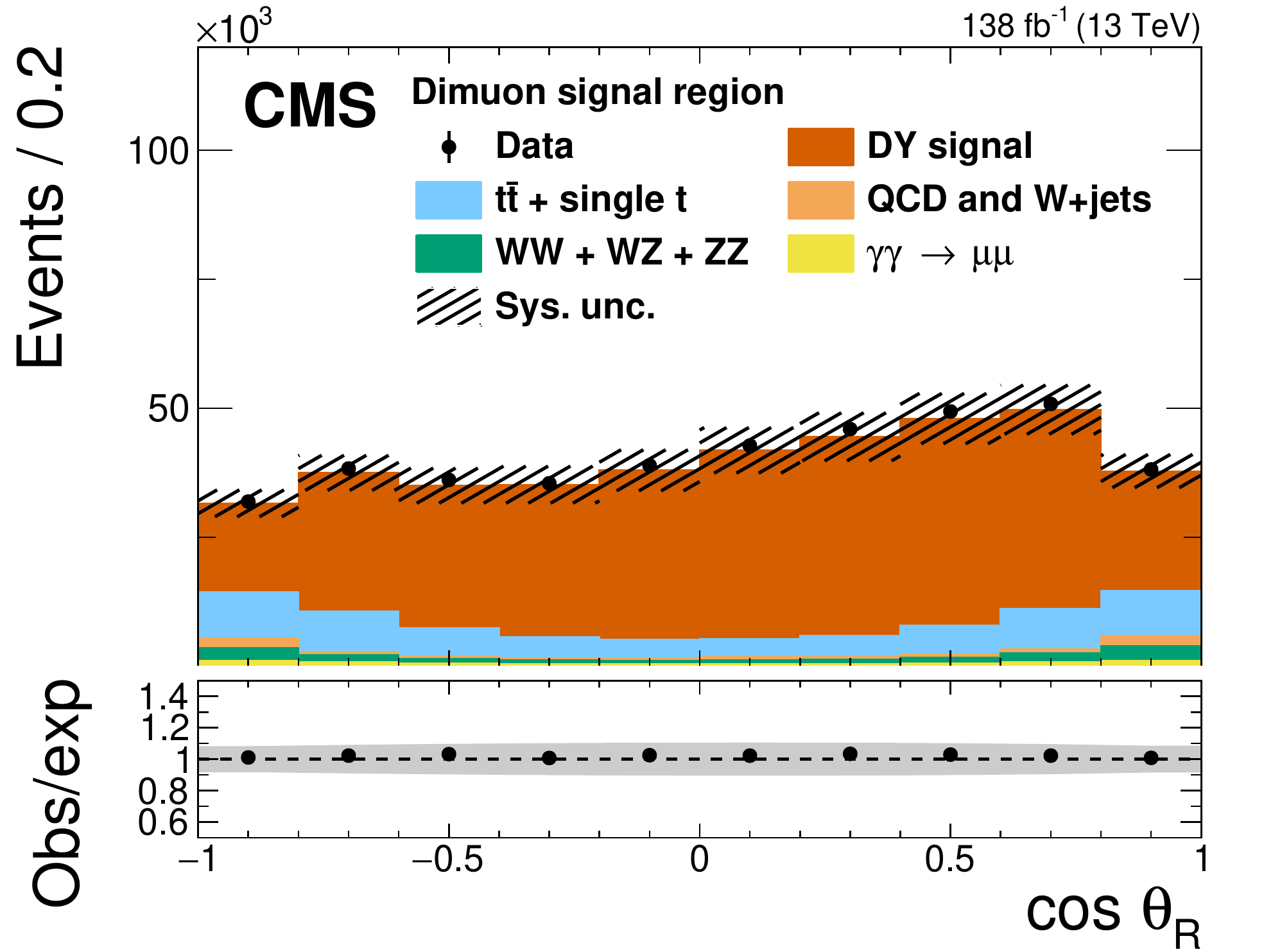}
  \includegraphics[width = 0.49 \textwidth]{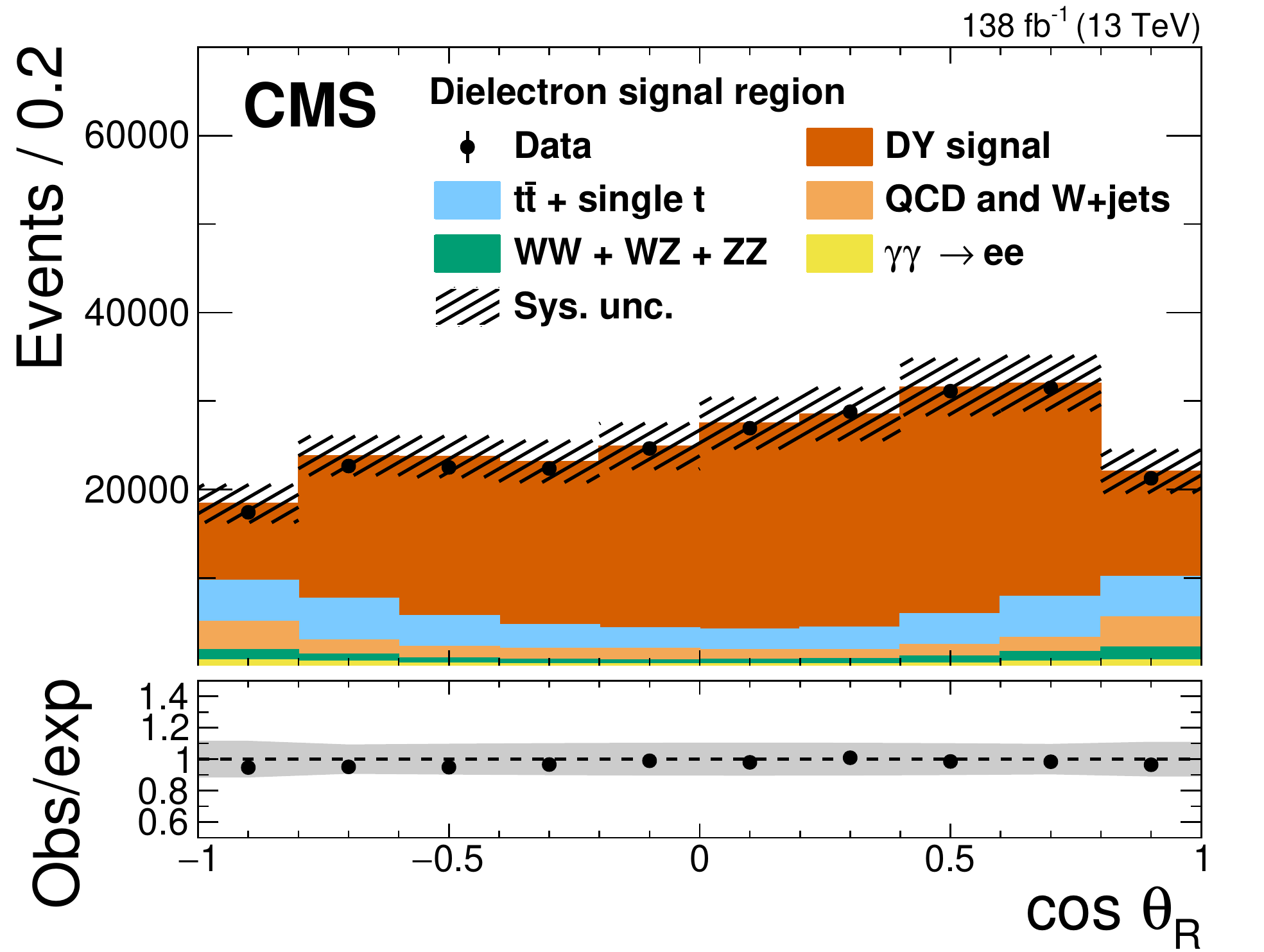}
  \includegraphics[width = 0.49 \textwidth]{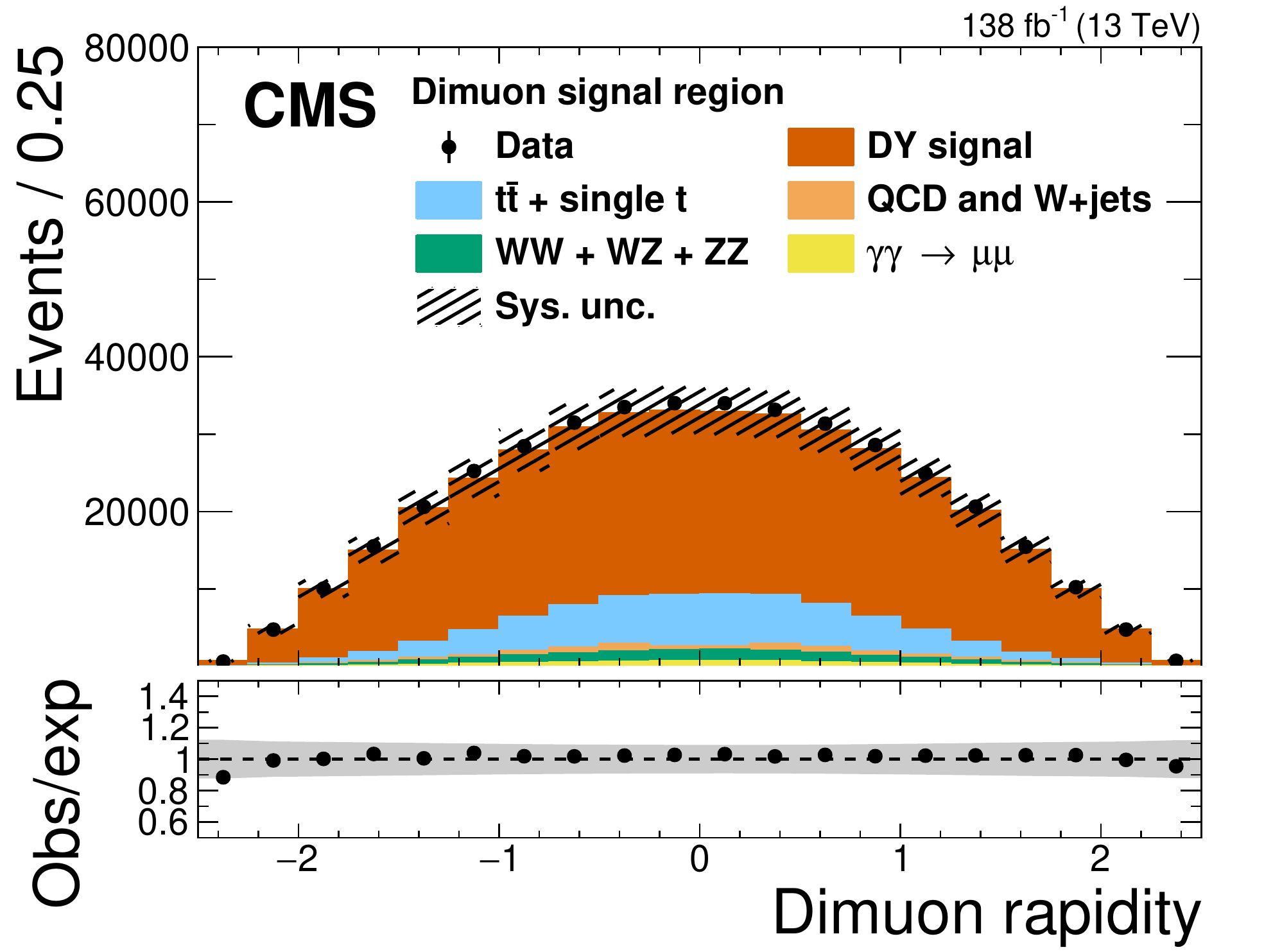}
  \includegraphics[width = 0.49 \textwidth]{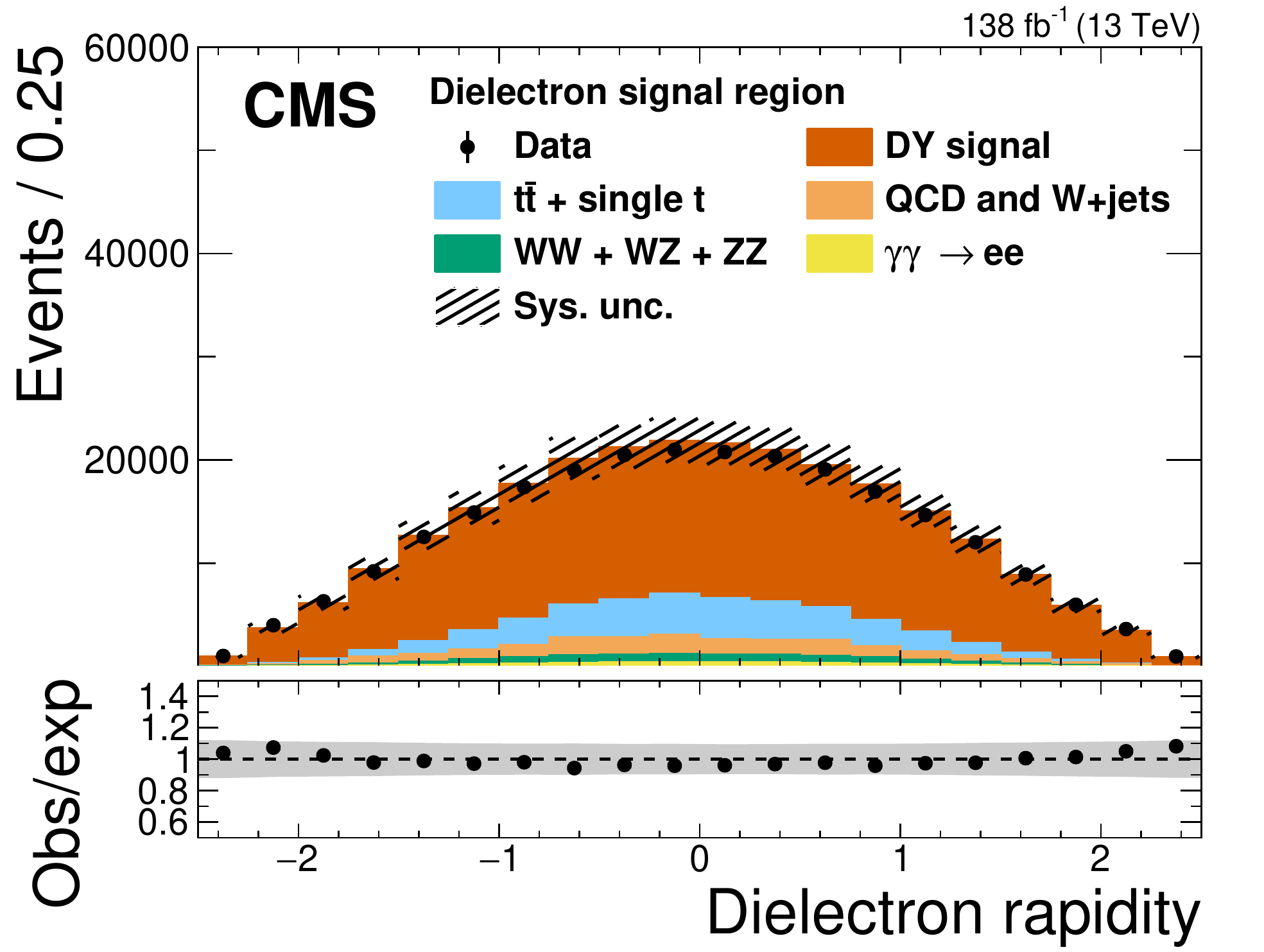}
  \caption{ A comparison of CMS data and expected signal and background distributions in dilepton invariant mass (upper row), \costhetar (middle row) and dilepton rapidity (lower row).
    The left plot shows the $\Pgm\Pgm$ channel and the right plot the $\Pe\Pe$ channel.
    The black points with error bars represent the data and their statistical uncertainties, whereas the combined signal and background expectation is shown as stacked histograms.
    The hatched band shows the systematic uncertainty in the expected signal and background yield. The sources of this uncertainty are discussed in Section~\ref{sec:sys_uncs}.
    The lower panels show the ratio of the data to the expectation. The gray bands represents the total uncertainty in the predicted yield.}
  \label{fig:Prefit_cmp}
\end{figure}

Figure~\ref{fig:Prefit_cmp} shows a comparison between measured $\Pgm\Pgm$ and $\Pe\Pe$ events and our estimates in the signal region after all corrections have been applied.
Good agreement is observed between the simulated and observed amount of $\Pgm\Pgm$ and $\Pe\Pe$ events.

\section{Template construction} \label{sec:temp_construct}

From the MC sample of DY events, templates for each of the three terms in Eq.~(\ref{eq:ang_xsec}) need to be constructed.
To avoid any of the templates having a negative yield in any of the bins, two reparameterizations are performed.
The angular distribution is rewritten in a slightly different form than Eq.~(\ref{eq:ang_xsec}): 
\begin{equation}
  \label{eq:ang_xsec_reparam}
  \frac{\rd\sigma}{\rd{}c_*} \propto \frac{3}{4(2+\alpha)}\biggl[ 1 + c_{*}^2 + \alpha (1 - c_{*}^2) + \frac{4 (2 + \alpha)}{3} \; \AFB \; c_{*} \biggr].
\end{equation}
This form is equivalent to Eq.~(\ref{eq:ang_xsec}) for $\alpha = 2 \Aze/(2-\Aze)$.
The term corresponding to NLO QCD production ($1 - c_{*}^2$) is now strictly positive.

The three templates, $f_\mathrm{S}$, $f_{\alpha}$, and $f_\mathrm{A}$, can now be constructed to represent the $1 + c_{*}^2$, $1 - c_{*}^2$, and $c_*$ 
terms in Eq.~(\ref{eq:ang_xsec_reparam}). 
These templates are constructed by reweighting MC events using generator-level quantities but are binned in the reconstructed variables.
Note that the generator-level $c_*$ and the reconstructed \csr can have different signs.
To minimize the impact of statistical fluctuations in our MC simulations, each event is used twice in the template construction. 
Each event is used once with $+c_*$ and once with $-c_*$, and each use is given half weight to keep the normalization unchanged.
The following reweighting factors are used to construct the $f_\mathrm{S}$, $f_{\alpha}$, and $f_\mathrm{A}$ templates respectively:
\begin{equation}
  \label{eq:sym_reweight}
  w_\mathrm{S}(\abs{c_*}) = \frac{1 + c_*^2}{1+c_*^2+\alpha\left(1-c_*^2\right)},
\end{equation}
\begin{equation}
  \label{eq:alpha_reweight}
  w_{\alpha}(\abs{c_*}) = \frac{1 - c_*^2}{1+c_*^2+\alpha\left(1-c_*^2\right)},
\end{equation}
\begin{equation}
  \label{eq:asym_reweight}
  w_\mathrm{A}(c_*) = \frac{c_*}{1+c_*^2+\alpha\left(1-c_*^2\right)}.
\end{equation}

The $\alpha$ values in the denominator are extracted from fits to generator level distributions of our MC simulation events. 

Then a histogram of CMS data events, $f_\text{data}$, could be parameterized by:
\begin{equation}
  f_\text{data} = \sum_jf^j_\text{bkg} + N(\alpha) \left[\alpha \; f_{\alpha} + f_\mathrm{S} + \frac{\AFB}{N(\alpha)} f_\mathrm{A} \right],
  \label{eq:template_fit_v1}
\end{equation}
where the normalization factor $N(\alpha) = 3/[4(2+\alpha)]$ has been introduced for convenience and $f^j_\text{bkg}$ represents templates for 
the different backgrounds.

To avoid the negative values of the antisymmetric template ($f_\mathrm{A}$), 
two positive-valued linear combinations of $f_\mathrm{A}$ and $f_\mathrm{S}$, are constructed as:
\begin{equation}
  f_\mathrm{S} + \frac{\AFB}{N(\alpha)} f_\mathrm{A} = (1 + \frac{\AFB}{N(\alpha)})\frac{f_\mathrm{S} + f_\mathrm{A}}{2} + (1 - \frac{\AFB}{N(\alpha)})\frac{f_\mathrm{S} - f_\mathrm{A}}{2} 
  \equiv (1 + \frac{\AFB}{N(\alpha)})f_{+} + (1 - \frac{\AFB}{N(\alpha)})f_{-}.
  \label{eq:template_redef}
\end{equation}
Then CMS data in each mass bin can be fit by the following parameterization:
\begin{equation}
  f_\text{data} = \sum_jf^j_\text{bkg} + N(\alpha) \left[ \alpha \; f_{\alpha} + (1 + \frac{\AFB}{N(\alpha)})f_{+} + (1 - \frac{\AFB}{N(\alpha)})f_{-}\right],
  \label{eq:template_fit}
\end{equation}
where the asymmetry \AFB and $\alpha$ are allowed to float.

Separate templates are made for seven different mass bins to facilitate the extraction of the \AFB and \Aze coefficients as functions of mass.
The bin edges used (in units of \GeV) are: 170, 200, 250, 320, 510, 700, 1000 and infinity,
where the last bin contains all events with dilepton mass greater than 1000\GeV.

Within each mass bin, the two-dimensional (2D) templates are binned in \csr and lepton pair rapidity $\abs{y}$.
Because of the acceptance effects, there are very few events with high rapidity and large $\abs{\csr}$.
Therefore, a different \csr binning is adopted at high rapidity.
There are four bins in $\abs{y}$ with bin edges at: 0.0, 0.6, 1.0, 1.5, and 2.4. 
For $\abs{y}<1.0$, there are eight uniform \csr bins from -1.0 to 1.0. 
For $\abs{y} > 1.0$, the pairs of bins with $\abs{\csr} > 0.5$ are merged leaving six \csr bins in this region.
Additionally, there are few events with large rapidity at high mass so for the highest two mass bins (700--1000 and $>$1000\GeV)
the two rapidity bins with $\abs{y} > 1.0$ are merged leaving three $\abs{y}$ bins in this region.

There are several templates for various background processes. 
Because of their similar shapes and the small magnitude of the $\PQt\PW$ background, the top quark related background processes (\ttbar, $\PQt\PW$) are combined 
into a single background template.
Diboson backgrounds ($ \PW\PW$, $\PZ\PZ$, $\PW\PZ$) are combined into a single background template as well.
The $\PGg \PGg \to \Pell \Pell$ process is a separate background template and constitutes 1--4\% of the total cross section in the mass range of the measurement.
The MisID estimate is also included as a separate background template.
A $\PZ/\PGg^* \to \PGt\PGt$ template is included as well.
To minimize statistical fluctuations, the templates for the top and MisID backgrounds are symmetrized in \costhetar. 
The templates for the diboson, $\PZ/\PGg^* \to \PGt\PGt$, and $\PGg \PGg \to \Pell \Pell$ processes are not symmetrized because 
these processes have perceptible asymmetries. 

Differential measurements of \AFB and \Aze are performed by fitting each mass bin independently, with a set of nuisance parameters uncorrelated with those in other mass bins. 
The results of an inclusive measurement in which all mass bins ($m > 170\GeV$) are fit simultaneously is also reported.
Within each mass bin all three years are fit simultaneously as separate categories. 
Two sets of fits are performed to extract \AFB and \Aze: one in which the \AFB and \Aze parameters in the $\Pe\Pe$ channel and $\Pgm\Pgm$ channels are allowed 
to float independently, and one in which they are fit with common \AFB and \Aze parameters.
An additional set of fits is performed to extract $\Delta\AFB$ and $\Delta\Aze$. 
In these fits, the \AFB and \Aze parameters in the $\Pe\Pe$ channel are used as a reference 
and the \AFB and \Aze parameters in the $\Pgm\Pgm$ channel are constructed in relation to them using the definitions of $\Delta\AFB$ and $\Delta\Aze$.
The \AFB and \Aze values in the $\Pe\Pe$ channel are allowed to freely float in the measurements of $\Delta\AFB$ and $\Delta\Aze$.
These measurements of $\Delta\AFB$ and $\Delta\Aze$ are performed separately for each mass bin, as well as inclusively across all mass bins.

\section{Systematic uncertainties}\label{sec:sys_uncs}

Systematic uncertainties in the normalization and shape of templates arise from a variety of sources 
and are defined through nuisance parameters in the likelihood. 
For systematic uncertainties that can change the shape of a template, shifted templates are constructed by varying 
the source of the systematic uncertainty up and down within its uncertainty. 
Shape uncertainties are incorporated into the likelihood by interpolation between the nominal and shifted templates, constrained with a Gaussian prior. 
The interpolation is calculated with a sixth-order polynomial for shifts smaller than one Gaussian $\sigma$, 
and with a linear function for shifts beyond one Gaussian $\sigma$. 
Normalization uncertainties are included using log-normal priors.

\textit{PDFs}:
variations in PDFs can change the dilution factor and kinematic distributions of the MC simulation samples. 
The 100 NNPDF set replicas are converted to 60 Hessian eigenvector variations~\cite{Carrazza:2015aoa}. 
For each of these 60 variations an ``up'' template is constructed using the variation's weight. 
A ``down'' template is then constructed that symmetrizes the deviation from the nominal template. 
Each variation is treated as a separate shape uncertainty yielding 60 nuisance parameters associated with PDFs in the fit.

\textit{MisID background estimate}:
the uncertainty in the normalization of the MisID background and the uncertainty in the shape of the MisID \costhetar distribution are considered separately.
Their normalizations are assigned a 50\% uncertainty based on closure studies of the misidentification rate technique performed in simulation.
The uncertainties in the shape corrections come from the statistical uncertainties of the same-sign validation and systematic uncertainties 
reflecting the shape differences between the same-sign and opposite-sign MisID estimates.

\textit{Lepton efficiencies}:
scale factors are derived and applied to the simulated MC events to account for the differences
in the detector performance between CMS data and the MC simulation. The efficiencies for the trigger,
lepton identification, and lepton isolation are measured as functions of lepton \pt and $\eta$ using the ``tag-and-probe''~\cite{Sirunyan:2018, CMS:2020uim} method for both data and simulation.
For muons, separate uncertainties are included for the muon trigger, identification, and isolation efficiencies, with the uncertainty in reconstruction efficiency being negligible. 
For electrons, isolation is part of the identification requirements rather than a separate selection, 
so separate uncertainties are included for the electron trigger, reconstruction, and identification/isolation efficiencies. 
The efficiency uncertainties also include the effects of timing issues in the ECAL endcaps and muon chambers that caused 
inefficiencies in the first-level trigger in the range of 0--3\% for the different data-taking periods.
For each of these uncertainties, up and down templates are constructed from MC simulations using the up and down values from the uncertainty in the scale factors.

\textit{Lepton scale corrections}:
the muon momentum and electron energy scales can be affected by detector alignment and imperfect calibration.
Such issues are corrected by additional energy and momentum scale corrections applied to both simulation and data.
A bias in the muon momentum reconstruction can occur because of the differences 
in the tracker alignment between CMS data and simulation, as well as a residual magnetic
field mismodeling. 
The muon momentum scale corrections are applied using the procedure described
in Refs.~\cite{Bodek:2012id,CMS:2019ied}. 
The electron energy deposits, as measured in the ECAL, are subject to a set of
corrections involving information from both the ECAL and tracker~\cite{CMS:2020uim}.
Separate up and down templates are constructed for the electron smearing, and muon and electron scales
based on the uncertainty of these corrections.

\textit{Cross sections}:
a systematic uncertainty is attributed to the normalization of signal and background samples estimated by MC event generators.
Based on NNLO calculations using the \FEWZ v3.1 simulation code~\cite{Gavin:2010az},
a 3\% uncertainty is assigned to the DY cross section.
Based on NLO calculations, a 5\% uncertainty is attributed to the backgrounds from top quark decays~\cite{Czakon:2011xx}, 
and a 4\% uncertainty in the diboson backgrounds~\cite{ Cascioli:2014yka, Gehrmann:2014fva, Campbell:2011bn}.
A 6\% uncertainty is attributed to the normalization of the $\PGg\PGg\to\ell\ell$ background, 
and it has been verified that this value covers the differences of samples generated
using Suri--Yennie structure functions and the LuxQED~\cite{Manohar:2016nzj,Manohar:2017eqh} photon PDFs.

\textit{Statistical uncertainties in templates}:
we treat the uncertainty due to the finite number of MC events, as well as the finite number of data events used to build the MisID templates,
using a modified ``Barlow--Beeston--like'' approach~\cite{Conway:2011in}, 
which adds a Gaussian uncertainty reflecting the number of events accumulated in each template bin.
However, template bins at $\pm\costhetar$ have partially correlated uncertainties
because of events being used multiple times in the construction of signal, MisID background, and top quark background templates, but not reused
in the $\PGg\PGg \to \Pell\Pell$, $\PGt\PGt$, and diboson background templates.
To account for this correlation, 
additional constraint terms are added in the fit.
These additional terms constrain the differences in the Barlow--Beeston--like nuisance parameters in correlated bins 
with Gaussian constraints.
For each channel and mass bin, the correlation coefficient between bins at $\pm\costhetar$ is computed and the standard deviation of the Gaussian function chosen 
conservatively based on the maximum amount of uncorrelated uncertainty in each mass bin.
Based on the computed correlations, a variance of $0.1$ is used for the Gaussian constraints on the template bins in the first four mass bins, and 
a variance of $0.6$ for the Gaussian constraints on template bins in the highest three mass bins. 

\textit{Integrated luminosity}:
the integrated luminosities of the 2016, 2017, and 2018 data-taking periods are individually known with uncertainties in the 
1.2--2.5\% range~\cite{CMS:2021xjt, CMS:2018elu, CMS:2019jhq}, while the total 2016--2018 integrated luminosity has an uncertainty of 1.6\%, the improvement in precision reflecting the (uncorrelated) time evolution of some systematic effects.

\textit{Renormalization + factorization scale and strong coupling}:
the renormalization (\muR) and factorization scales (\muF), and strong coupling (\alpS) used in MC generation can also have an effect on the 
shape and normalization of the templates.
The \muR and \muF uncertainties are included by reweighting simulated events to match alternative scenarios where
\muR and \muF are scaled up and down by a factor of two, both independently and simultaneously.
The unphysical combinations where \muR and \muF are scaled in opposite directions are not included.
Three nuisance parameters, describing \muR, \muF and combined \muR and \muF scale variations are included in the fit and are taken to be uncorrelated with respect to each other.
Simulated events are also reweighted to account for variations of \alpS and used to construct up and down templates.
A central value of $\alpS = 0.118$ and variations of $\pm$ 0.0015 are used~\cite{Butterworth:2015oua}.

\textit{Background shape corrections}:
as discussed in Section~\ref{sec:backgrounds}, a shape correction based on $\Pe\Pgm$ data events is applied to MC simulation derived backgrounds from top quarks and diboson events. 
The uncertainties in this shape correction are modeled with four independent nuisance parameters. 
Each parameter allows the values of the shape correction in a particular $\abs{\costhetar}$ bin to vary up and down within its statistical uncertainty.

\textit{DY \pt correction}:
it is known that the DY dilepton \pt spectrum is difficult to model at low \pt~\cite{ATLAS:2019zci,CMS:2019raw}. 
Because acceptance is correlated with \pt, mismodeling of the \pt spectrum can result in mismodeling of the DY contribution. 
For this reason, the DY MC \pt spectrum in each mass bin is corrected based on data.
For each mass bin used in the measurement, the normalized DY MC simulation \pt distribution is compared with the normalized data distribution, less the contributions from backgrounds. 
A correction factor is then derived based on the ratio of these two distributions.
When building the templates used in the measurement, events are reweighted with this additional factor 
so that the events used to build the templates match the \pt spectrum of data events.
To model the uncertainty in this correction seven separate nuisance parameters are used, one for each \pt bin.
For each parameter the correction in a particular \pt bin is varied up and down within its statistical uncertainty to produce up and down templates. 
The total uncertainty in these corrections leads to an uncertainty in the signal templates of roughly 1\%. 

\textit{Pileup}:
to account for the uncertainty originating from the differences in the measured and simulated pileup distribution, 
the total inelastic cross section is varied by $\pm$ 4.6\%~\cite{Sirunyan:2018nqx}, and the reweighting factor applied to MC simulated samples is recomputed. 
Up and down MC simulation templates are constructed based on the up and down variations of the reweighting factor. 

\textit{\PQb tag veto efficiency}:
scale factors dependent on jet flavor, \pt, and $\eta$ are applied to simulated
events to correct for differences in \PQb tagging efficiency and mistag rates between data and
simulation~\cite{BTV-16-002}. Up and down templates are constructed based on the uncertainty in these efficiencies.

\ptmiss \textit{modeling}:
the uncertainty in the calibration of the jet energy scale and resolution affects the modeling of \ptmiss.
These effects are estimated by changing the jet energy in the simulation up and down by one standard
deviation.
These up and down variations are used to construct sets of up and down templates for use in the fitting
procedure.

As discussed in Section~\ref{sec:analysis_strategy}, there is also an uncertainty in the final measurement of \AFB based
on the extrapolation from the fiducial region to the full phase space. 
This uncertainty is added as an additional source of systematic uncertainty when the correction is applied after the template fit.

The contributions from different sources of systematic uncertainty are evaluated by redoing the fit while different groups of nuisances are fixed to their nominal values
and taking the quadrature difference between the resulting uncertainty and the full uncertainty. 
A comparison of the magnitude of the different systematic uncertainties on the \AFB and $\Delta\AFB$ measurements is shown in Table~\ref{tab:sys_unc_mags}. 
The single largest source of systematic uncertainty in the measurement of \AFB are the PDFs, but it is a negligible source of uncertainty in the measurement of $\Delta\AFB$.
The dominant sources of systematic uncertainty are common to the electron and muon channels and thus their combination reduces the statistical uncertainty of 
the measurement but does not significantly reduce the systematic uncertainty.

\begin{table}
  \centering
  \captionof{table}{ A comparison of the magnitude of the different sources of systematic uncertainty for the measurement of \AFB when combining the muon and electron channels
    and for the measurement of $\Delta\AFB$. 
    Results for the 170--200\GeV mass bin are shown because that is the mass bin in which the systematic uncertainty 
    has the largest contribution to the total uncertainty; the results for other mass bins are similar. 
    Results are also reported as a fraction of the overall systematic uncertainty for the measurement of \AFB and $\Delta\AFB$.
    Sources are listed in order of the size of their contribution to the uncertainty in \AFB.}

    \label{tab:sys_unc_mags}
    \cmsTable{
      \begin{tabular}{ l c m{33mm} c m{33mm} }
        Source          & Unc. on \AFB ($\times10^{-3}$)  & \centering Frac. of total \newline sys. unc. on \AFB  
                        & Unc. on $\Delta\AFB$ ($\times10^{-3}$)   & \centering Frac. of total \newline sys. unc. on $\Delta\AFB$ \tabularnewline
        \hline
        PDFs                                        & 8.1                & \centering 47\%                       & 0.8                         & \centering 1\% \tabularnewline
        MC and MisID backgrounds stat. unc.         & 4.1                & \centering 12\%                       & 6.8                         & \centering 42\% \tabularnewline
        \alpS + \muF/\muR scales                    & 3.3                & \centering 8\%                        & 3.2                         & \centering 9\% \tabularnewline
        DY cross section                            & 3.0                & \centering 7\%                        & 0.9                         & \centering 1\% \tabularnewline
        Pileup                                      & 2.8                & \centering 5\%                        & 3.8                         & \centering 13\% \tabularnewline
        Fiducial correction                         & 2.7                & \centering 5\%                        & $<$0.1                      & \centering $<$1\% \tabularnewline
        \ttbar cross section                        & 2.7                & \centering 5\%                        & 0.1                         & \centering $<$1\% \tabularnewline
        DY \pt correction                           & 2.1                & \centering 3\%                        & 0.8                         & \centering 1\% \tabularnewline
        $\Pe\Pgm$ shape corrections                 & 1.8                & \centering 2\%                        & 0.4                         & \centering $<$1\% \tabularnewline
        Integrated luminosity                       & 1.2                & \centering 1\%                        & 1.0                         & \centering 1\% \tabularnewline
        Electron identification/isolation           & 1.0                & \centering 1\%                        & 2.7                         & \centering 6\% \tabularnewline
        Electron MisID normalization                & 0.9                & \centering 1\%                        & 4.3                         & \centering 17\% \tabularnewline
        Electron MisID shape                        & 0.8                & \centering $<$1\%                     & 2.6                         & \centering 6\% \tabularnewline
        \PQb tagging uncertainty                       & 0.8                & \centering $<$1\%                     & 0.3                         & \centering $<$1\% \tabularnewline
        \ptmiss uncertainties                       & 0.7                & \centering $<$1\%                     & 0.5                         & \centering $<$1\% \tabularnewline
        Muon identification/isolation               & 0.6                & \centering $<$1\%                     & 1.2                         & \centering 1\% \tabularnewline
        Muon MisID shape                            & 0.5                & \centering $<$1\%                     & 0.6                         & \centering $<$1\% \tabularnewline
        $\PGg\PGg$ cross section                    & 0.4                & \centering $<$1\%                     & 0.6                         & \centering $<$1\% \tabularnewline
        Muon MisID normalization                    & 0.4                & \centering $<$1\%                     & 0.1                         & \centering $<$1\% \tabularnewline
        Electron trigger                            & 0.4                & \centering $<$1\%                     & 1.2                         & \centering 1\% \tabularnewline
        Diboson cross section                       & 0.2                & \centering $<$1\%                     & 0.1                         & \centering $<$1\% \tabularnewline
        Electron reconstruction                     & 0.2                & \centering $<$1\%                     & 0.7                         & \centering $<$1\% \tabularnewline
        Muon momentum scale                         & 0.1                & \centering $<$1\%                     & 0.1                         & \centering $<$1\% \tabularnewline
        Electron momentum scale                     & 0.1                & \centering $<$1\%                     & 0.1                         & \centering $<$1\% \tabularnewline
        Muon trigger                                & 0.1                & \centering $<$1\%                     & 0.1                         & \centering $<$1\%\tabularnewline
    \end{tabular}
  }
\end{table}

\section{Results}\label{sec:results}

\begin{figure}[hbt!]
  \centering
  \includegraphics[width = 0.7 \textwidth]{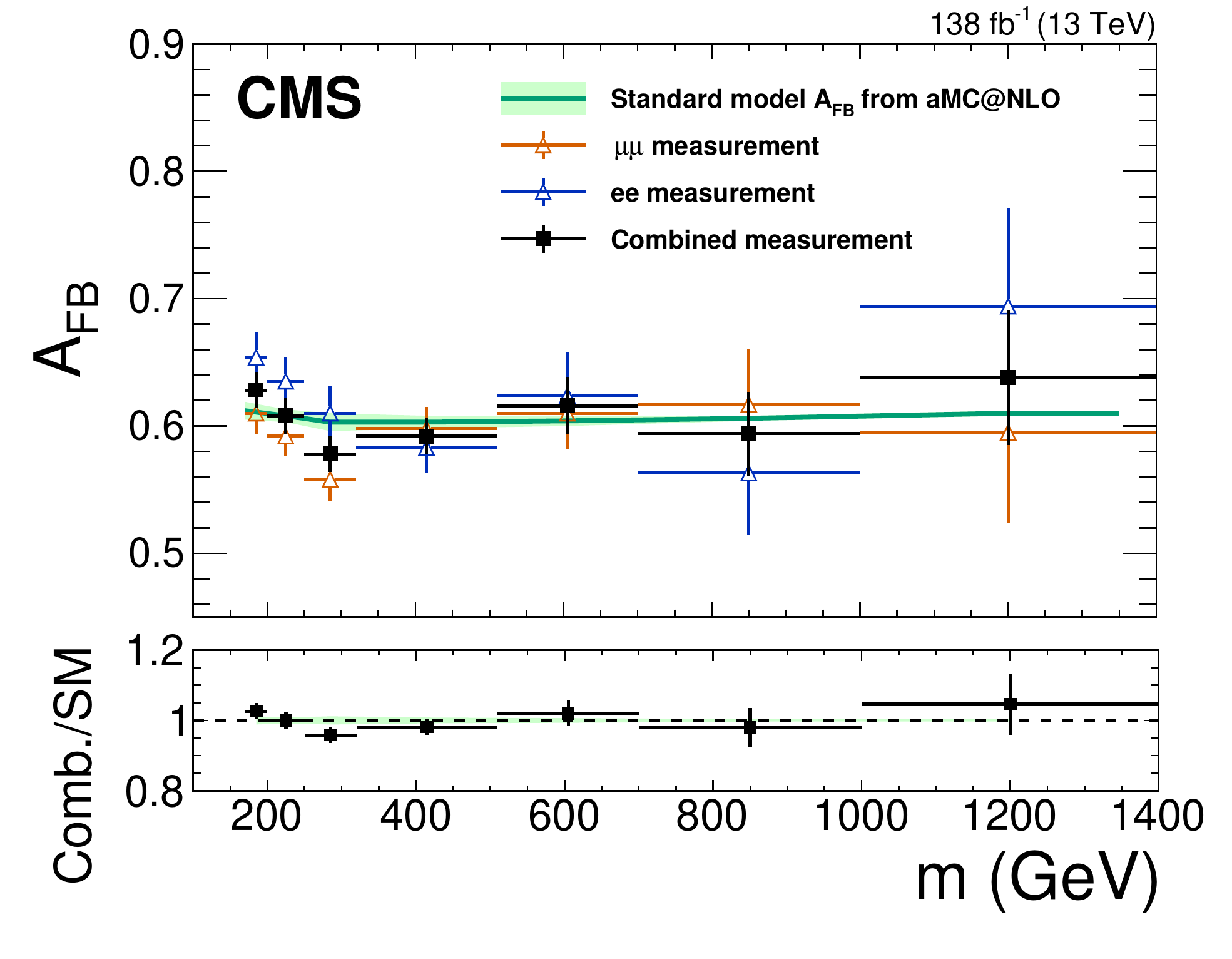}
  \caption{Measurement of the DY forward-backward asymmetry as a function the dilepton mass compared with the MC predictions. 
    The green line is the predicted value for \AFB from the \MGaMC simulation and the shaded green region its uncertainty.
    The red, blue, and black points and error bars represent the dimuon, dielectron, and combined measurements, respectively.
    Error bars on the measurements include both statistical and systematic components.
    The bottom panel shows the ratio between the combined measurement and the \MGaMC prediction. 
	In the bottom panel, the vertical error bars represent the uncertainty in the combined measurement and the shaded green band the uncertainty in the \MGaMC prediction.}
  \label{fig:AFB_mbins}
\end{figure}

\begin{figure}[hbt!]
  \centering
  \includegraphics[width = 0.7 \textwidth]{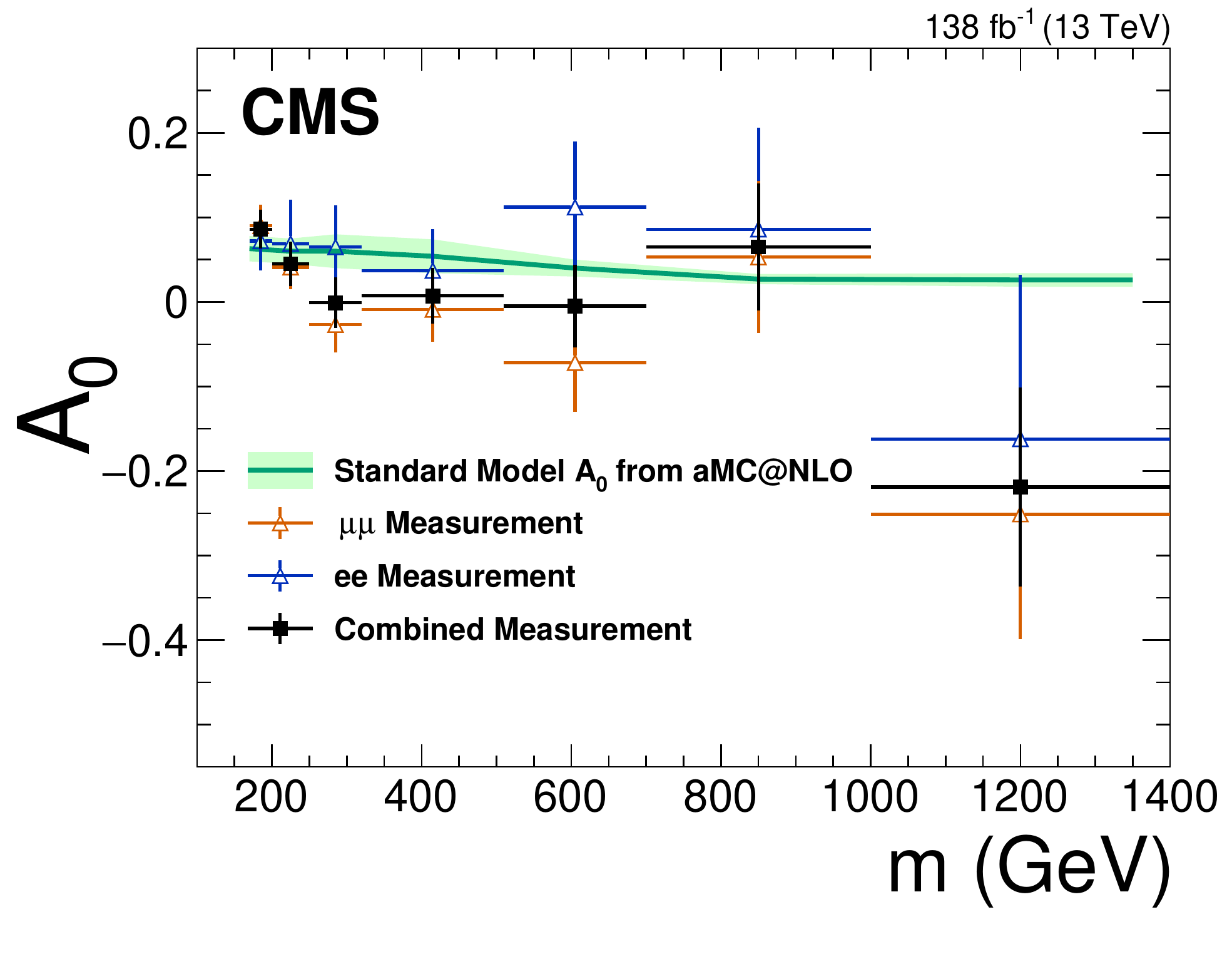}
  \caption{Measurement of the DY forward-backward asymmetry as a function the dilepton mass compared with the MC predictions. 
    The green line is the predicted value for \AFB from the \MGaMC simulation and the shaded green region its uncertainty.
    The red, blue, and black points and error bars represent the dimuon, dielectron, and combined measurements, respectively.
    Error bars on the measurements include both statistical and systematic components.}
  \label{fig:A0_mbins}
\end{figure}

\begin{figure}[hbt!]
  \centering
  \includegraphics[width = 0.65 \textwidth]{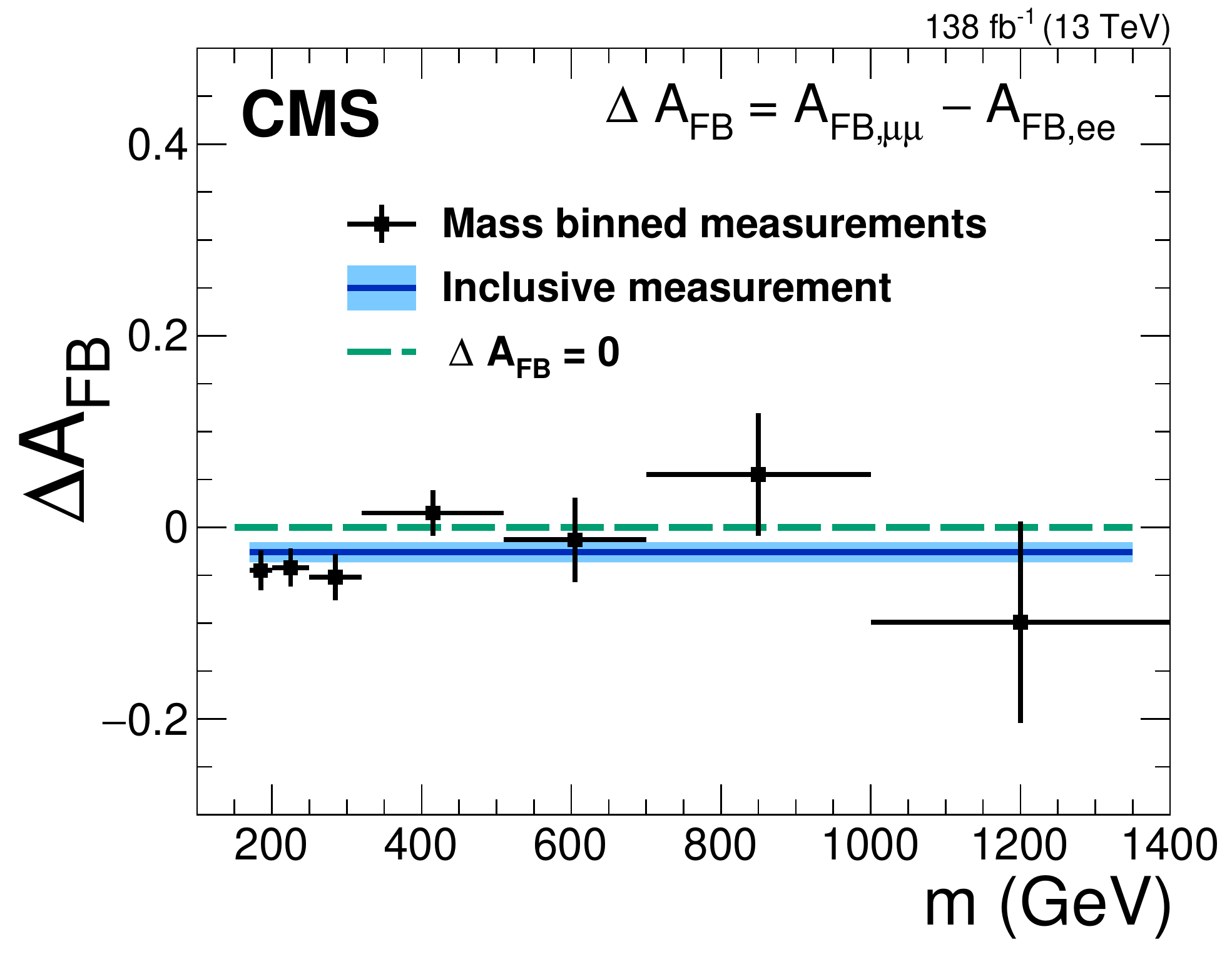}
  \caption{Measurement of the difference in forward-backward asymmetry between the dimuon and dielectron channels. 
    The green line is drawn at zero, the predicted value for $\Delta\AFB$ assuming lepton flavor universality.
    The black points and error bars represent the measurements of $\Delta\AFB$ in different mass bins.
    The blue line and shaded light blue region represent the inclusive measurement of $\Delta\AFB$ and corresponding uncertainty.
    The error bars on the measurements and the shaded region include both statistical and systematic components.}
  \label{fig:delta_AFB}
\end{figure}

\begin{figure}[htb!]
  \centering
  \includegraphics[width = 0.7 \textwidth]{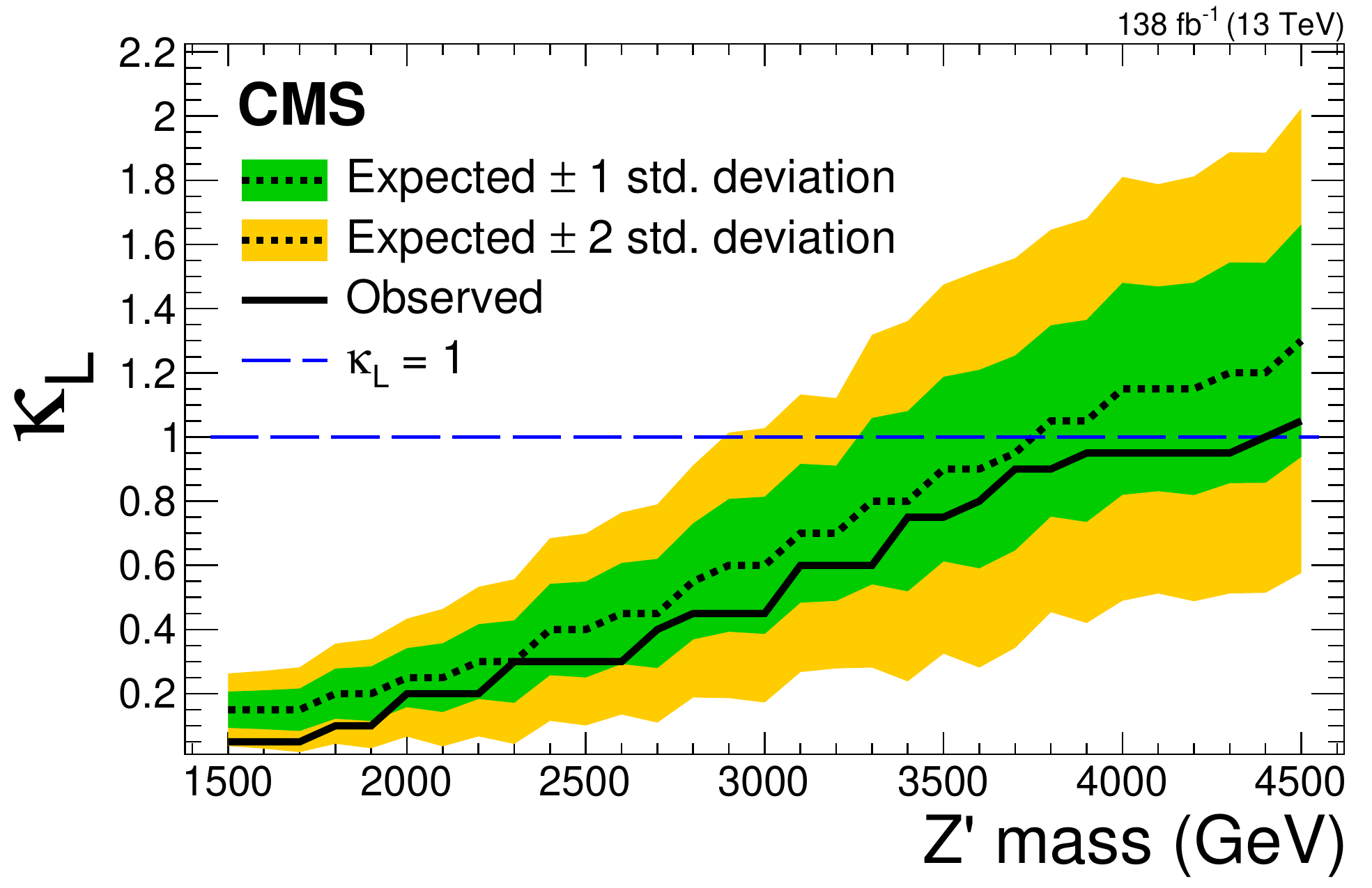}
  \captionof{figure}{
    Exclusion limits at 95\% \CL on the coupling parameter $\kappa_\mathrm{L}$ for a \PZpr in the sequential standard model as a function of the \PZpr mass. 
    The expected (observed) limit is shown by the dashed (solid) line.
    The inner and outer shaded areas around the expected limits show the 68\% (green) and 95\% (yellow) \CL intervals, respectively.
    The dashed blue line shows $\kappa_{L} = 1$ which corresponds to a \PZpr with exactly the same couplings as the SM \PZ boson.}
  \label{fig:Zp_lim}
\end{figure}

\begin{figure}[!hbtp]
  \begin{centering}
    \includegraphics[width = 0.49 \textwidth]{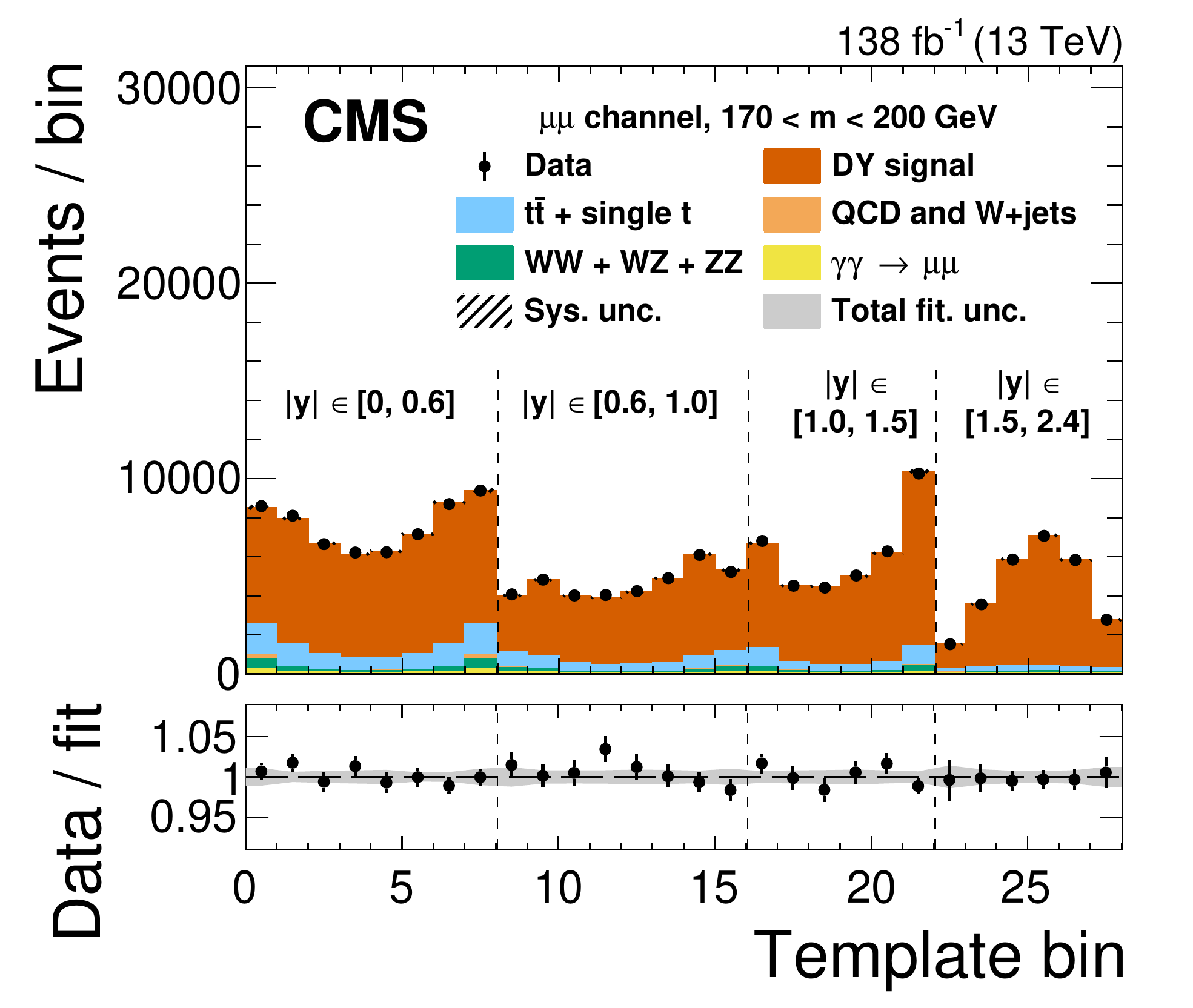}
    \includegraphics[width = 0.49 \textwidth]{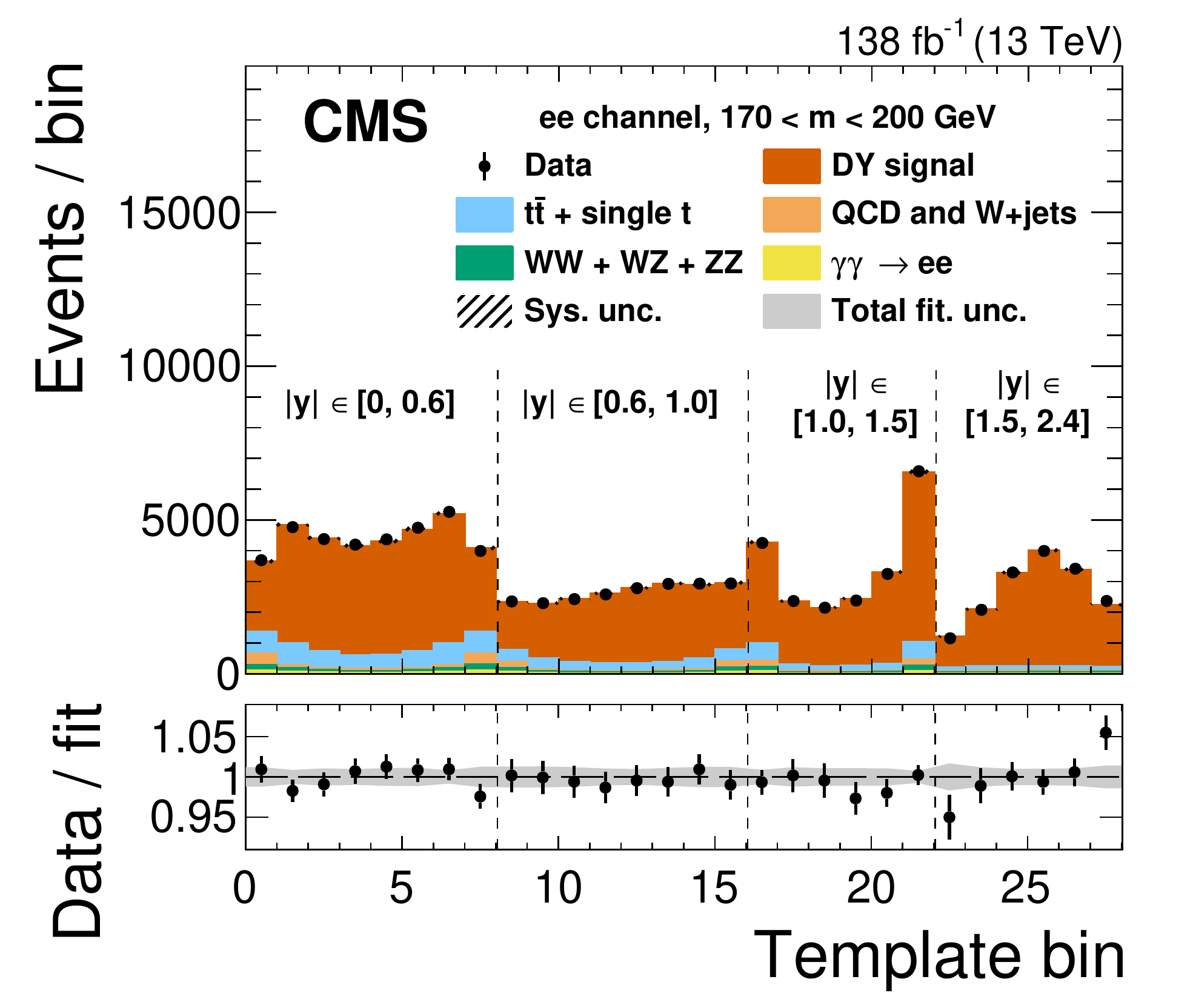}
    \includegraphics[width = 0.49 \textwidth]{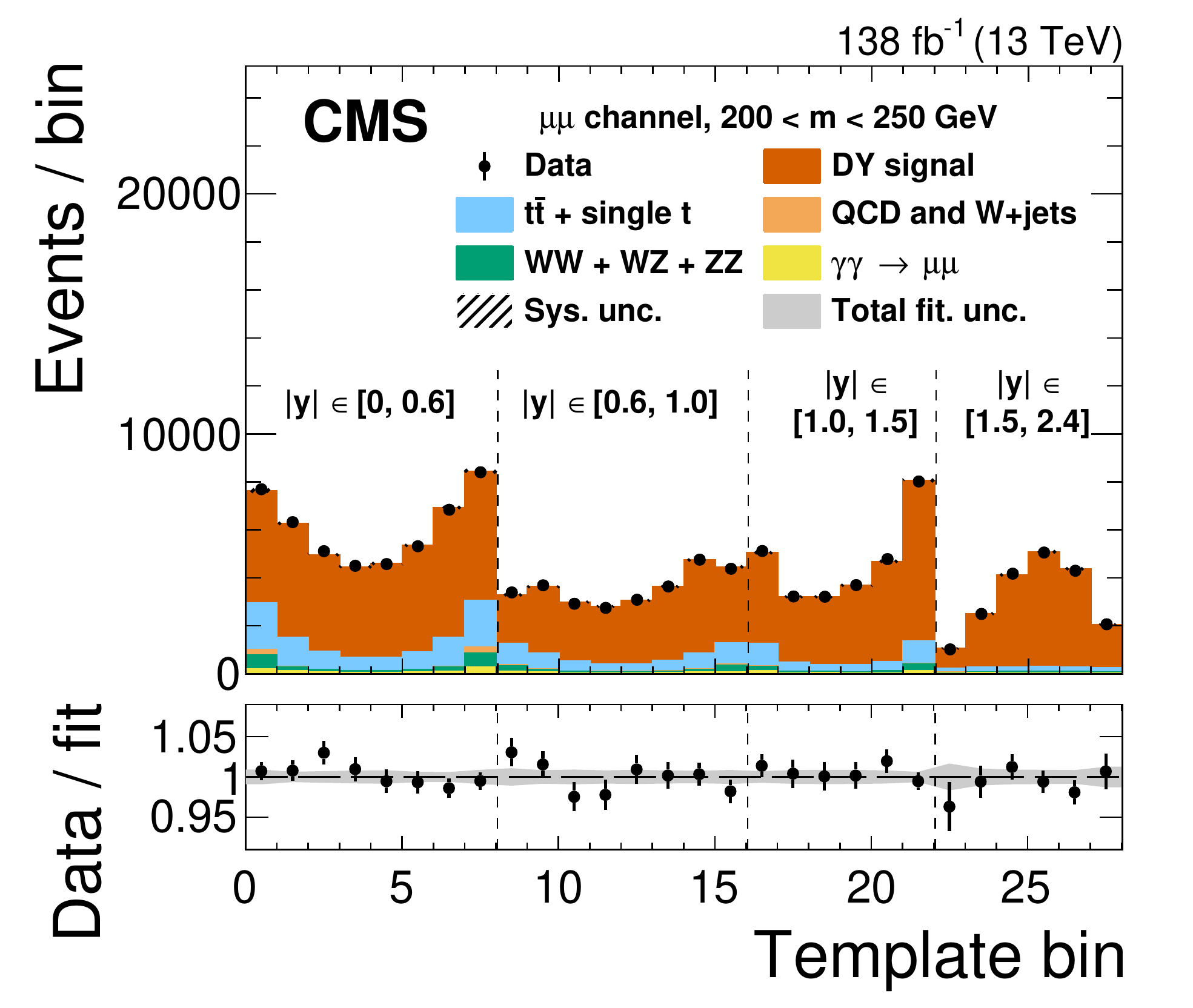}
    \includegraphics[width = 0.49 \textwidth]{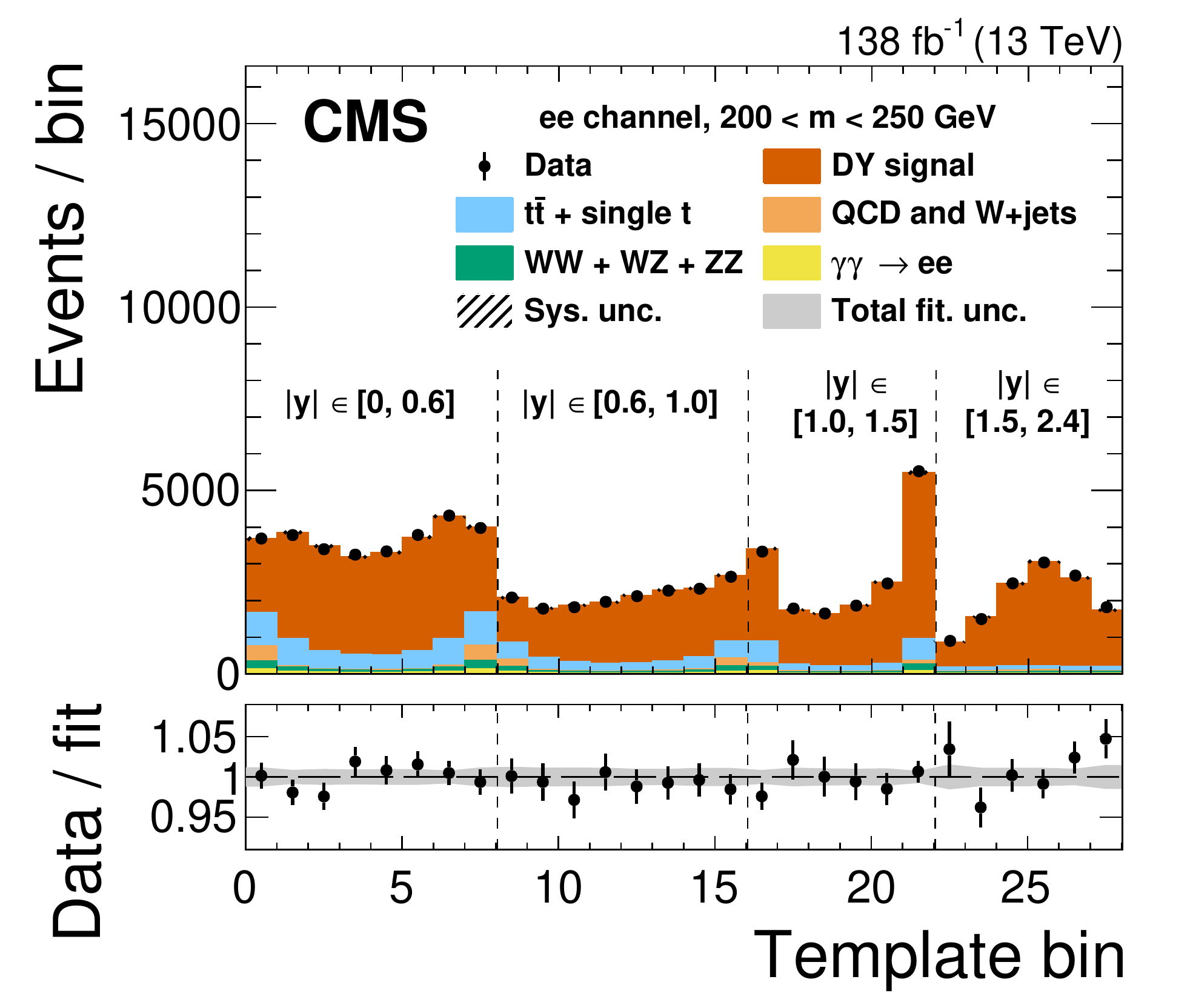}
    \includegraphics[width = 0.49 \textwidth]{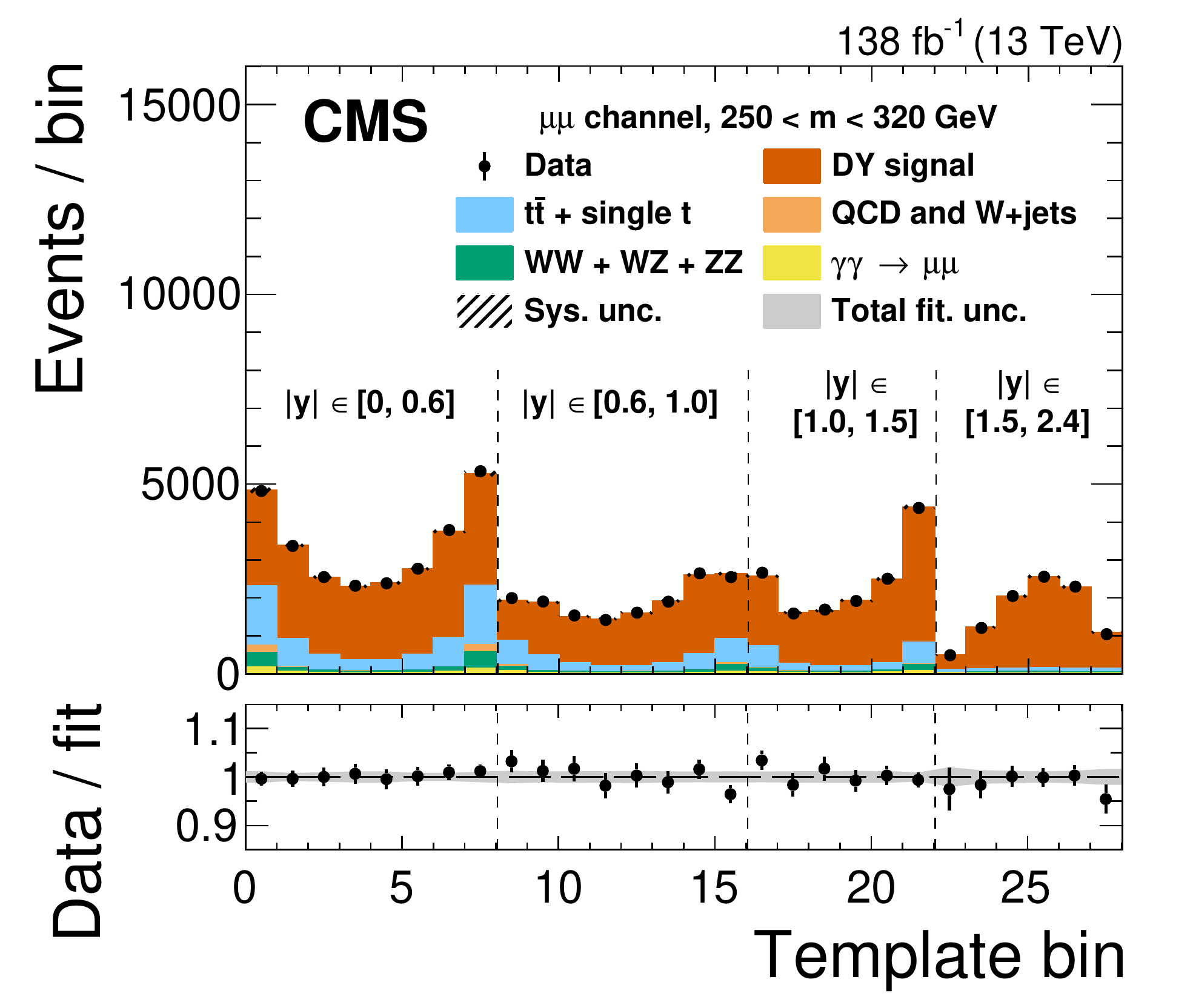}
    \includegraphics[width = 0.49 \textwidth]{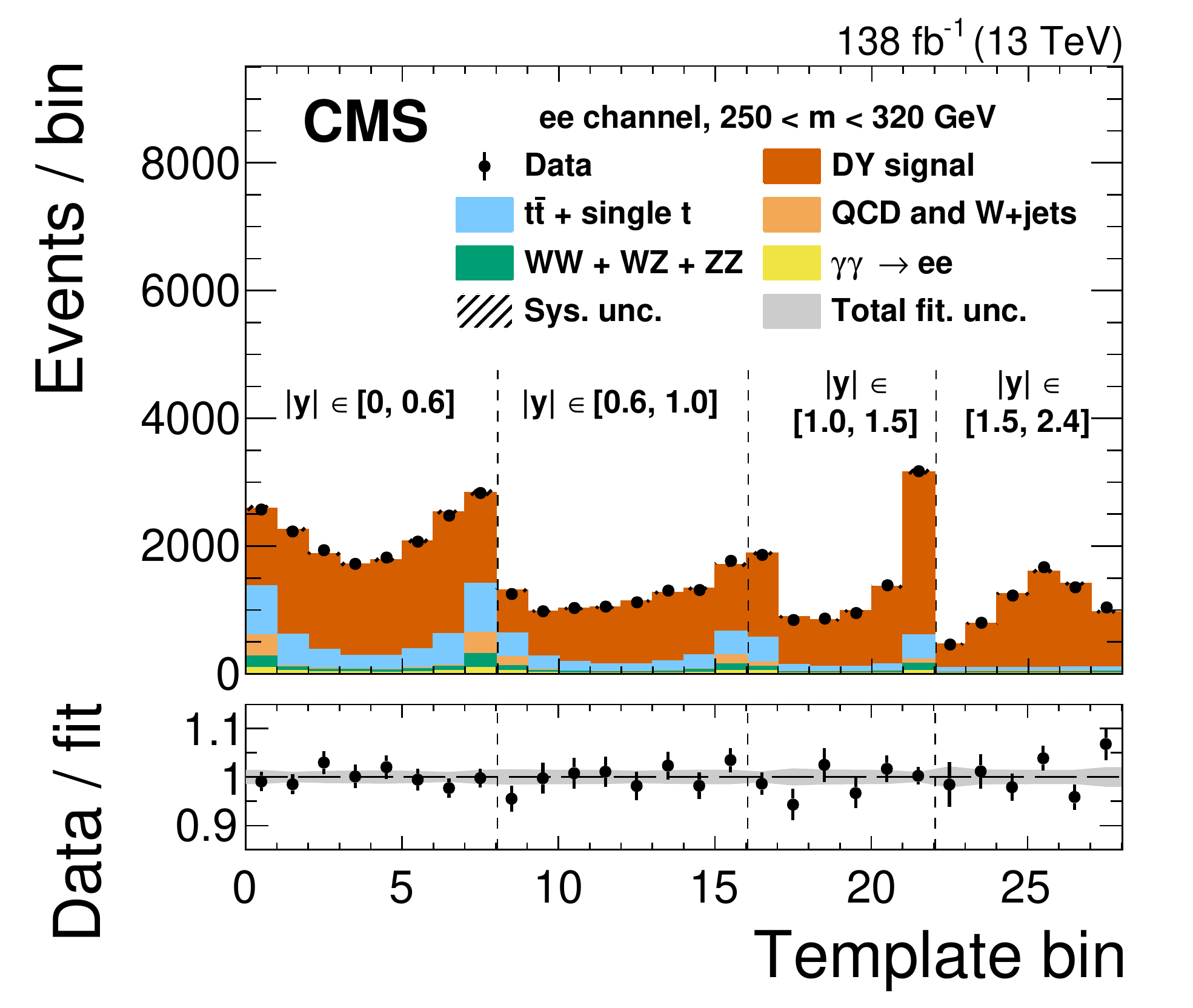}
  \end{centering}
  \caption{The postfit distributions in the 170--200, 200--250 and 250--320\GeV mass bins 
    are shown in upper, middle and lower rows, respectively.
    The left column is the $\Pgm\Pgm$ channel, and the right column the $\Pe\Pe$ channel.
    The contribution of the $\PGt\PGt$ background is not visible on the scale of these plots and has been omitted. 
    The 2D templates follow the \costhetar and $\abs{y}$ binning defined in Section~\ref{sec:temp_construct} but are presented here in one dimension, 
    with the dotted lines indicating the different $\abs{y}$ bins.
    The black points and error bars represent the data and their statistical uncertainties.
    The bottom panel in each figure shows the ratio between the number of events observed in data and the best fit value. 
    The gray shaded region in the bottom panel shows the total uncertainty in the best fit result.}
  \label{fig:postfit_fig1}

\end{figure}

\begin{figure}[!hbtp]
  \begin{centering}
    \includegraphics[width = 0.49 \textwidth]{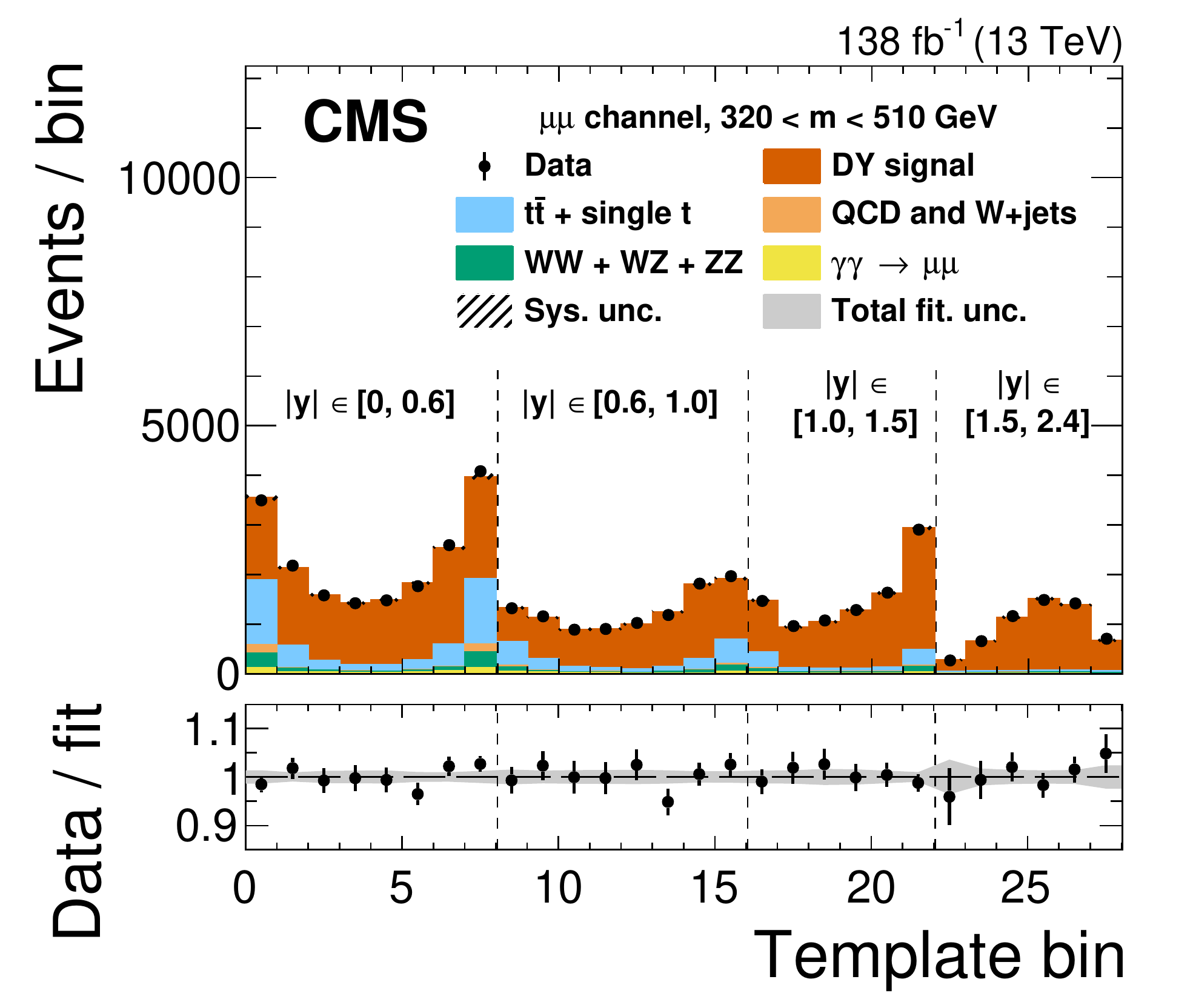}
    \includegraphics[width = 0.49 \textwidth]{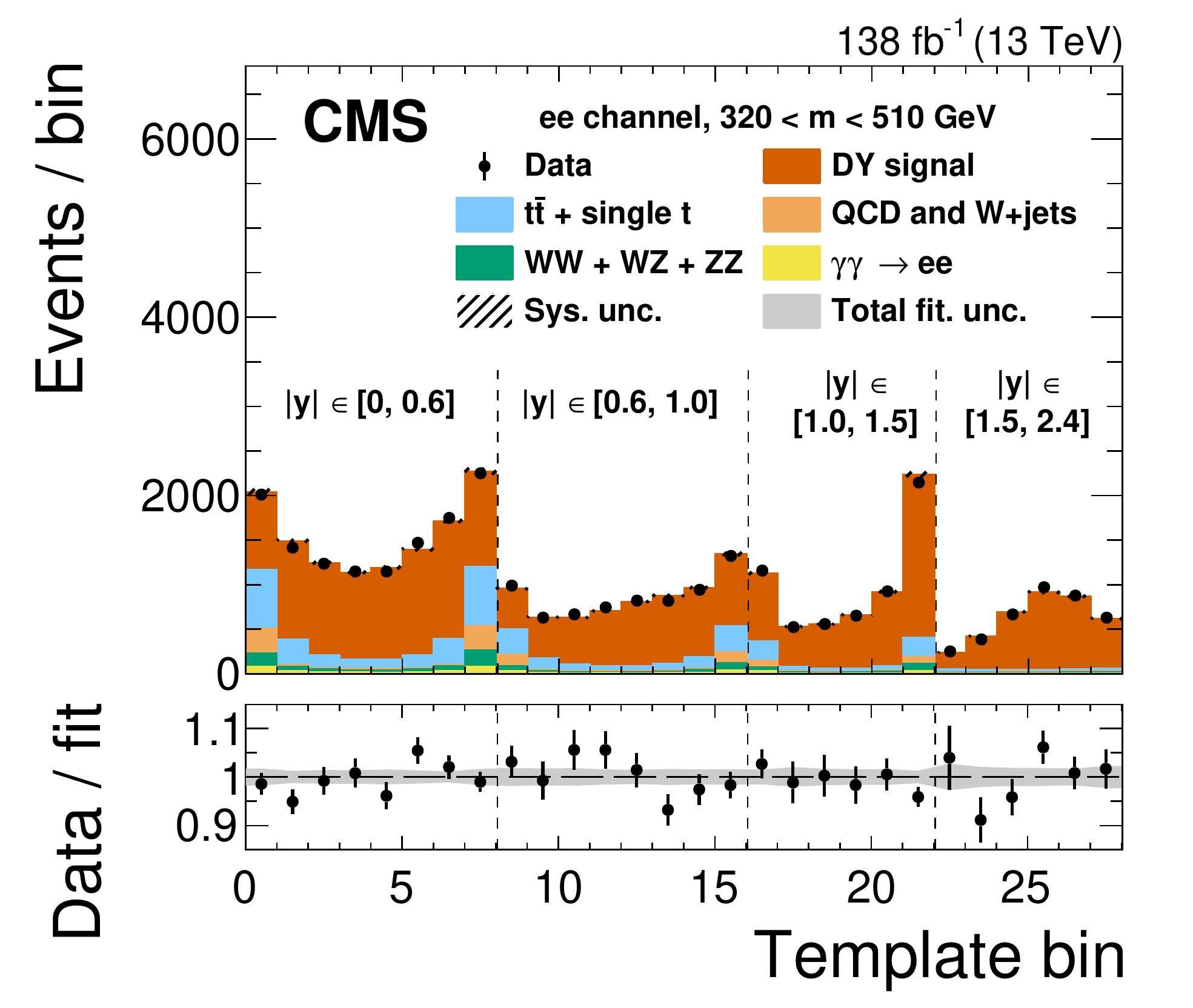}
    \includegraphics[width = 0.49 \textwidth]{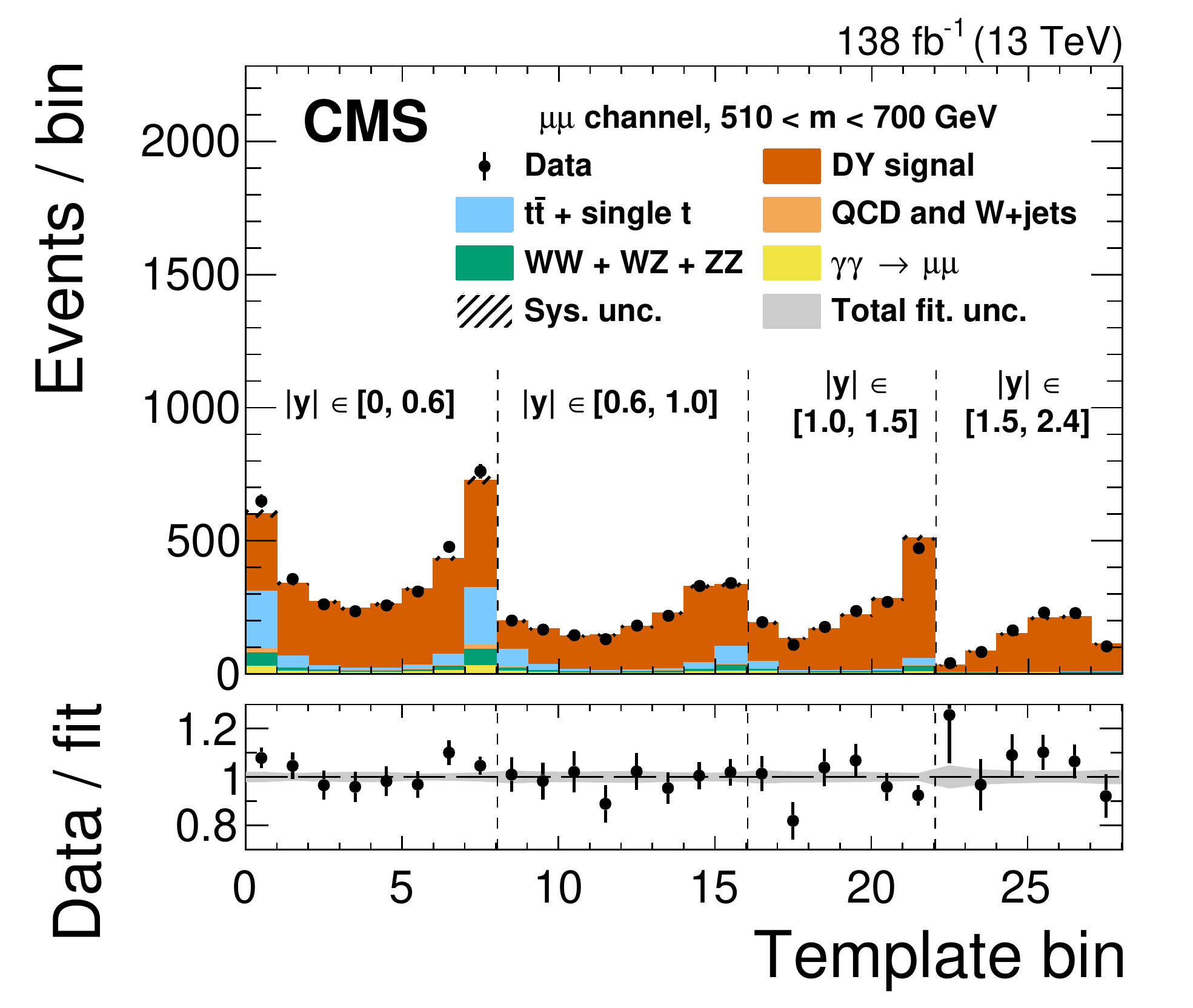}
    \includegraphics[width = 0.49 \textwidth]{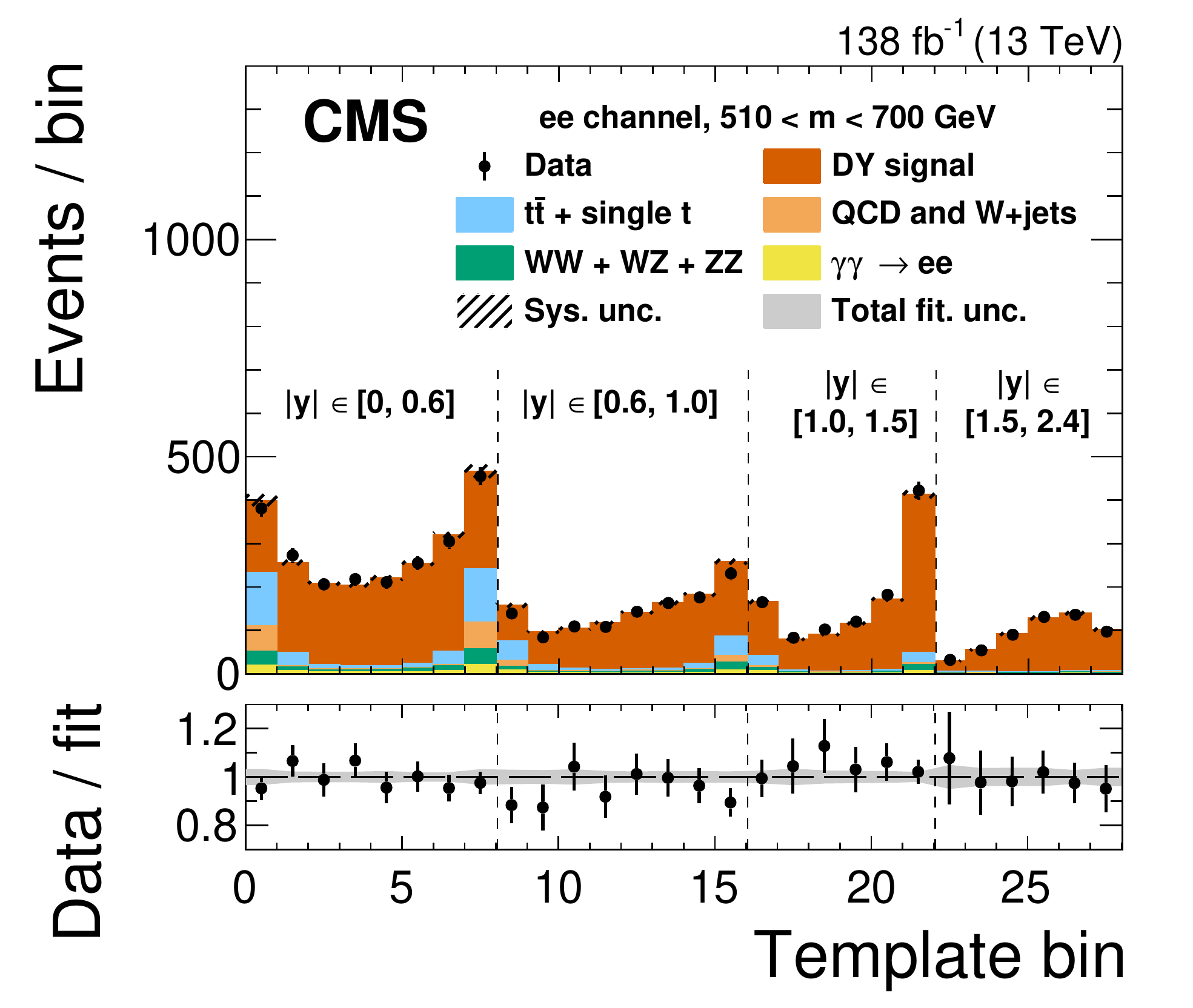}
  \end{centering}
  \caption{The postfit distributions in the 320--510 and 510--700\GeV mass bins are
    shown in the upper and lower rows, respectively. 
    The left column is the $\Pgm\Pgm$ channel, and the right column the $\Pe\Pe$ channel.
    The contribution of the $\PGt\PGt$ background is not visible on the scale of these plots and has been omitted. 
    The 2D templates follow the \costhetar and $\abs{y}$ binning defined in Section~\ref{sec:temp_construct} but are presented here in one dimension, 
    with the dotted lines indicating the different $\abs{y}$ bins.
    The black points and error bars represent the data and their statistical uncertainties.
    The bottom panel in each figure shows the ratio between the number of events observed in data and the best fit value. 
    The gray shaded region in the bottom panel shows the total uncertainty in the best fit result.}
  \label{fig:postfit_fig2}
\end{figure}

\begin{figure}[!hbtp]
  \begin{centering}
    \includegraphics[width = 0.49 \textwidth]{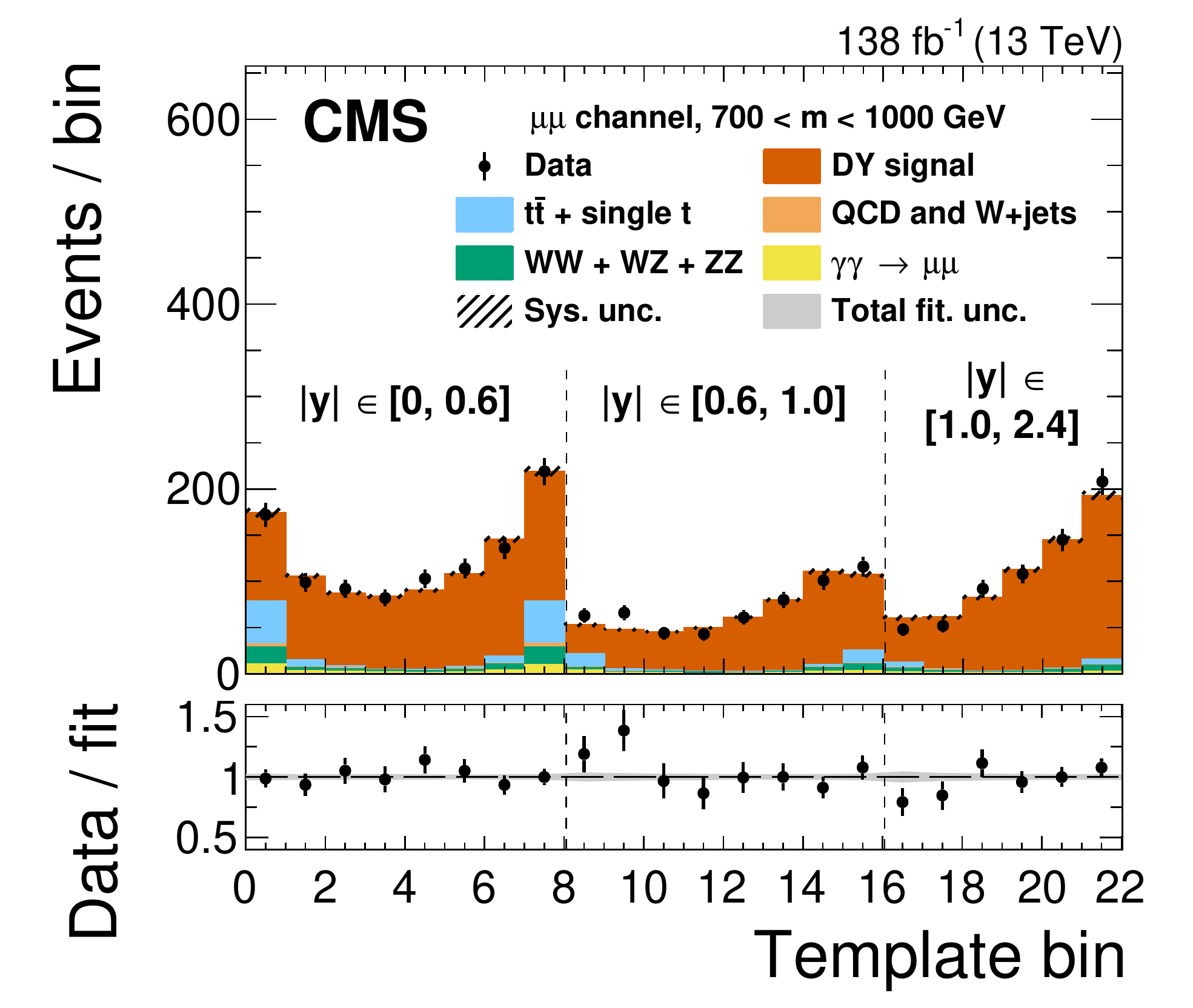}
    \includegraphics[width = 0.49 \textwidth]{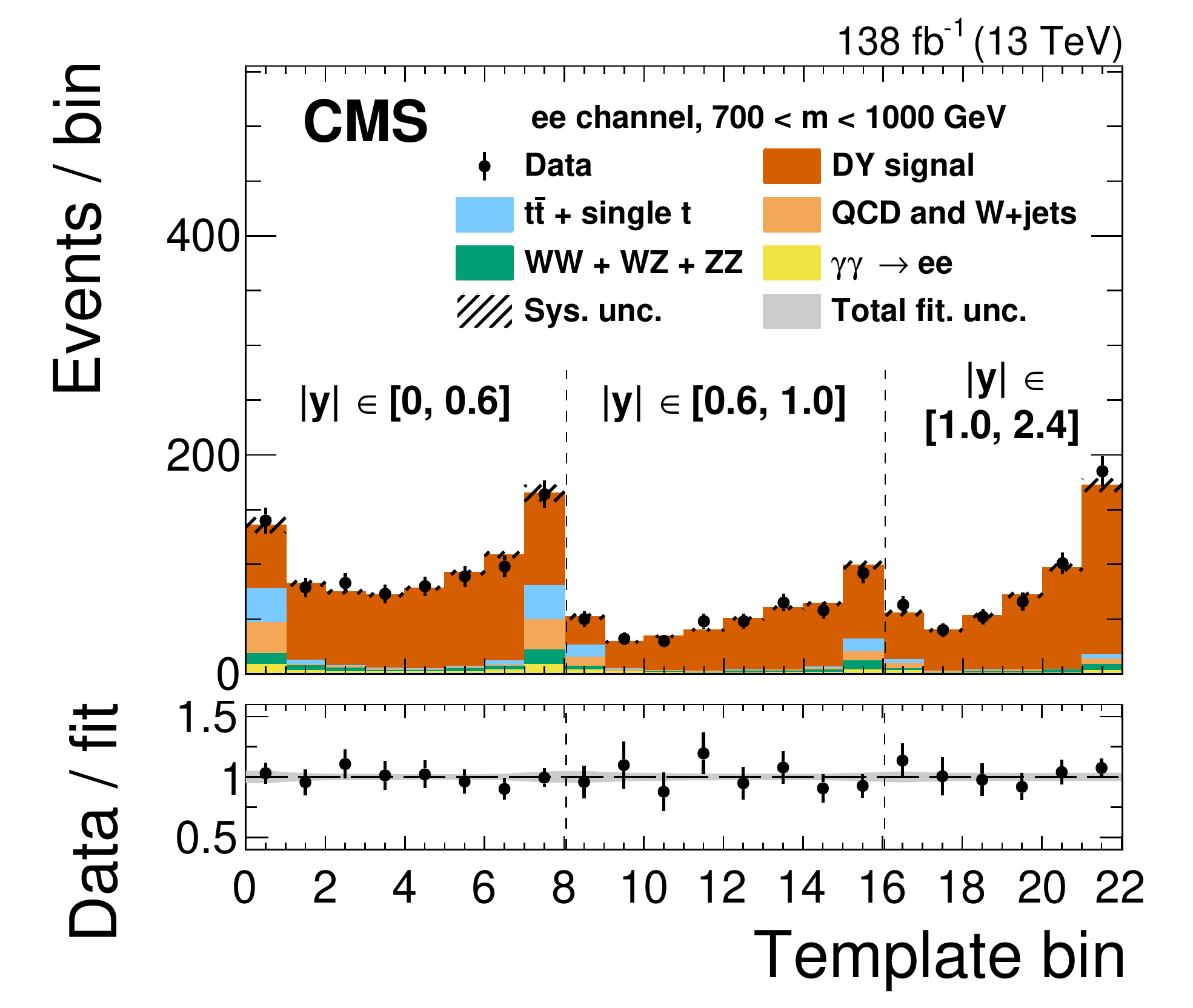}
    \includegraphics[width = 0.49 \textwidth]{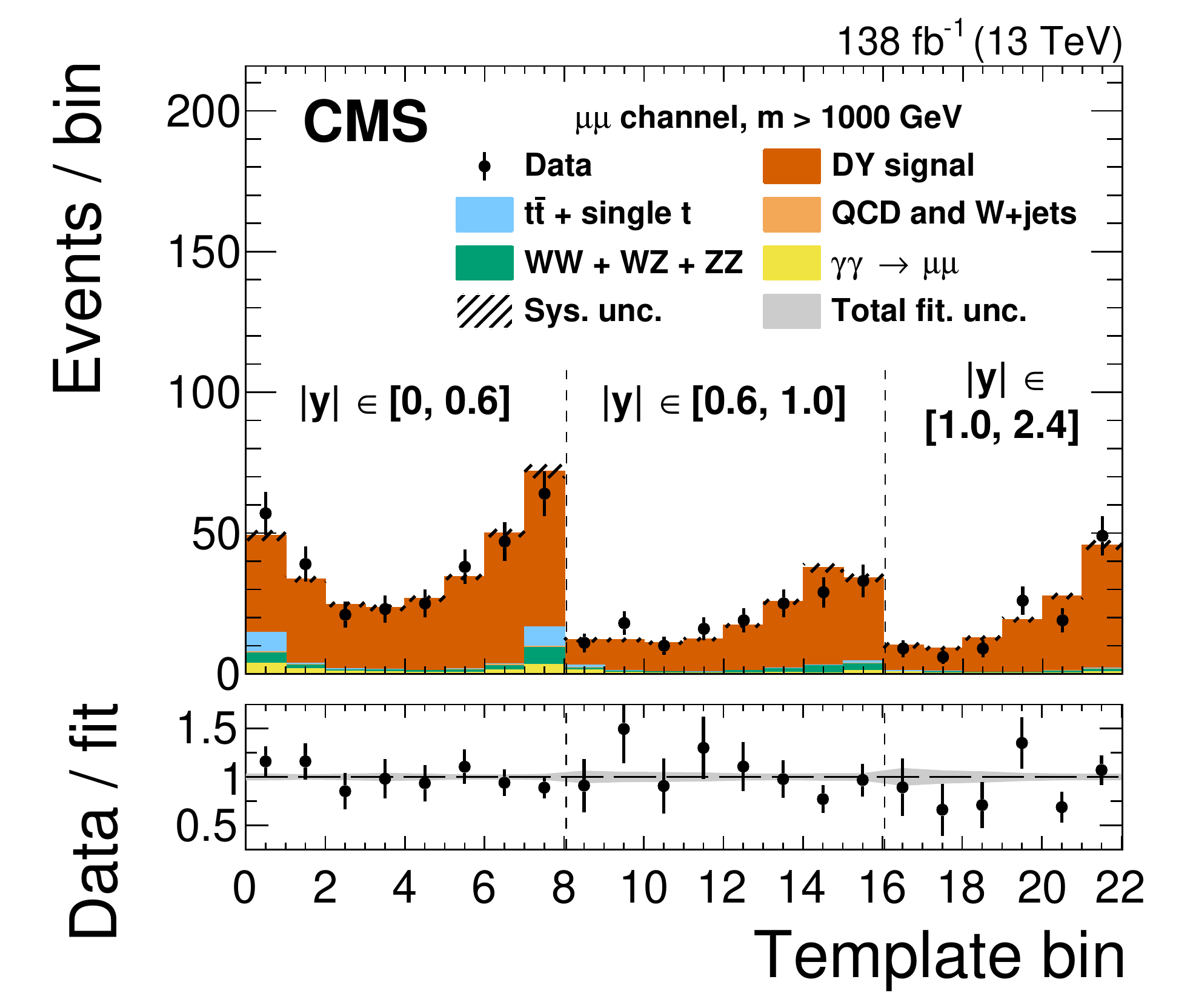}
    \includegraphics[width = 0.49 \textwidth]{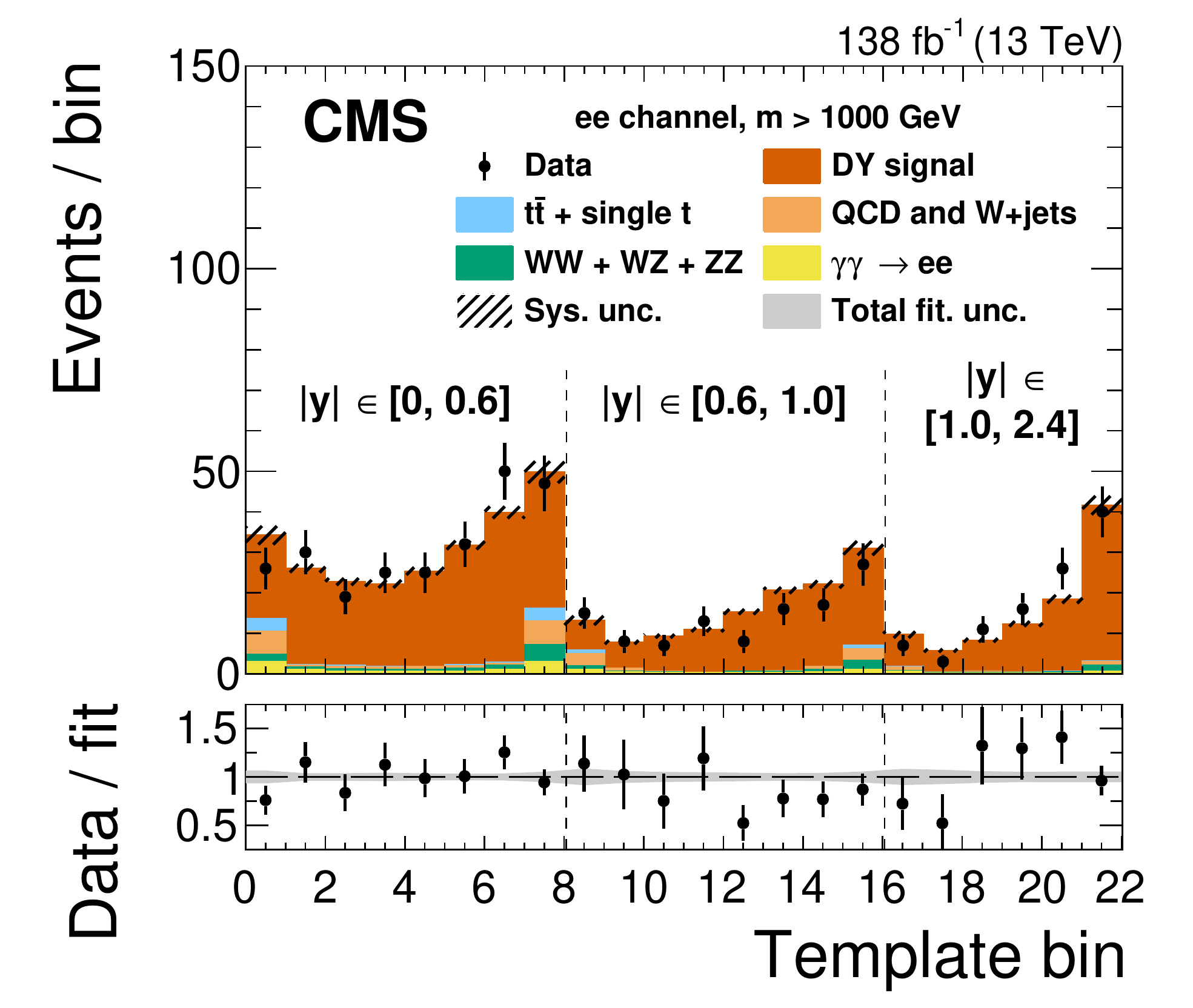}
  \end{centering}
  \caption{The postfit distributions in the 700--1000 and $>$1000\GeV mass bins are shown 
    in the upper and lower rows, respectively. 
    The left plot is the $\Pgm\Pgm$ channel, and the right plot the $\Pe\Pe$ channel.
    The contribution of the $\PGt\PGt$ background is not visible on the scale of these plots and has been omitted. 
    The 2D templates follow the \costhetar and $\abs{y}$ binning defined in Section~\ref{sec:temp_construct} but are presented here in one dimension, 
    with the dotted lines indicating the different $\abs{y}$ bins.
    The black points and error bars represent the data and their statistical uncertainties.
    The bottom panel in each figure shows the ratio between the number of events observed in data and the best fit value. 
    The gray shaded region in the bottom panel shows the total uncertainty in the best fit result.}
  \label{fig:postfit_fig3}
\end{figure}

For all relevant results presented in this section, the best-fit value of the parameter 
and an approximate 68\% \CL confidence interval are extracted following the procedure described in Section~3.2 of~\cite{Khachatryan:2014jba}.
The results for the template fits to data to extract \AFB in different mass bins are presented in Table~\ref{tab:afb_m_results} and shown graphically in Fig.~\ref{fig:AFB_mbins};
the results for \Aze are presented in Table~\ref{tab:a0_m_results} and shown graphically in Fig.~\ref{fig:A0_mbins}.
The results for the extraction of $\Delta\AFB$ and $\Delta\Aze$ in different mass bins and inclusively are presented in Table~\ref{tab:delta_afb_results}.
The results for $\Delta\AFB$ are also shown graphically in Fig.~\ref{fig:delta_AFB}.
The resulting exclusion limits are shown in Fig.~\ref{fig:Zp_lim}.
The contribution of $\PGg\PGg \to \Pell \Pell$ events as compared to DY events in each mass bin is shown in Table~\ref{tab:phot_ind_fracs}.
Comparisons of the data and the postfit predictions are shown in Figs.~\ref{fig:postfit_fig1}--\ref{fig:postfit_fig3}. 

\begin{table}[!htbp]
  \centering
  \topcaption{ Results for the measurement of \AFB from the maximum likelihood fit to data in different dilepton mass bins in the different channels
    as well as an inclusive measurement across all mass bins.
    The first and second uncertainties listed with each measurement are statistical and systematic, respectively.
    The last column lists the predictions of \AFB and associated uncertainties from \MGaMC.}
    \label{tab:afb_m_results}
    \cmsTable{
      \begin{tabular}{ c  c  c  c  c}
        Mass (\GeVns)  &  \AFB muons &  \AFB  electrons &   \AFB combined  & \MGaMC Pred. \\
        \hline

  		170--200        & $0.610 \pm 0.012  \pm 0.011$    & $0.654 \pm 0.015 \pm 0.012$  & $0.628 \pm 0.009  \pm 0.011$  & $0.612 \pm 0.007  $  \\ 
  		200--250        & $0.592 \pm 0.012  \pm 0.010$    & $0.635 \pm 0.015 \pm 0.011$  & $0.608 \pm 0.009  \pm 0.010$  & $0.608 \pm 0.006  $  \\ 
  		250--320        & $0.558 \pm 0.014  \pm 0.009$    & $0.610 \pm 0.018 \pm 0.010$  & $0.578 \pm 0.011  \pm 0.009$  & $0.603 \pm 0.007  $  \\ 
  		320--510        & $0.598 \pm 0.014  \pm 0.009$    & $0.583 \pm 0.018 \pm 0.009$  & $0.592 \pm 0.011  \pm 0.008$  & $0.603 \pm 0.005  $  \\ 
  		510--700        & $0.610 \pm 0.027  \pm 0.008$    & $0.624 \pm 0.033 \pm 0.009$  & $0.616 \pm 0.021  \pm 0.008$  & $0.604 \pm 0.004  $  \\ 
  		700--1000       & $0.617 \pm 0.042  \pm 0.009$    & $0.563 \pm 0.048 \pm 0.008$  & $0.594 \pm 0.032  \pm 0.008$  & $0.606 \pm 0.002  $  \\ 
  		$>$1000         & $0.595 \pm 0.070  \pm 0.011$    & $0.694 \pm 0.076 \pm 0.014$  & $0.638 \pm 0.052  \pm 0.011$  & $0.610 \pm 0.002  $  \\
        [\cmsTabSkip] 
        Inclusive, Mass $>$170  & $0.602 \pm 0.006  \pm 0.007$    & $0.628 \pm 0.008 \pm 0.007$  & $0.612 \pm 0.005  \pm 0.007$ & $0.608 \pm 0.006$    \\

    \end{tabular}
  }

\end{table}

\clearpage

\begin{table}[!htb]
  \centering
  \topcaption{ Results for the measurement of \Aze from the maximum likelihood fit to data in different dilepton mass bins in the different channels 
    as well as an inclusive measurement across all mass bins.
    The first and second uncertainties listed with each measurement are statistical and systematic, respectively.
    The last columns lists the predictions of \Aze and associated uncertainties from \MGaMC.
    To help in the interpretation of these results, we also list the average dilepton $\pt$ of the data events in each mass bin.}
    \label{tab:a0_m_results}
    \cmsTable{
      \begin{tabular}{ c c c c c c}
        Mass (\GeVns) & Avg. \pt (\GeVns) &  \Aze muons  & \Aze electrons  & \Aze combined  & \MGaMC Pred. \\
        \hline

        170--200     & 38  & $\ \ \ 0.090  \pm 0.010  \pm 0.023$   &$\ \ \ 0.072 \pm 0.013 \pm 0.032$ & $\ \ \ 0.086  \pm 0.008   \pm 0.022$  & $0.06 \pm 0.01$ \\ 
        200--250     & 43  & $\ \ \ 0.041  \pm 0.011  \pm 0.024$   &$\ \ \ 0.069 \pm 0.015 \pm 0.050$ & $\ \ \ 0.045  \pm 0.009   \pm 0.024$  & $0.06 \pm 0.01$ \\ 
        250--320     & 48  & $   -0.027  \pm 0.016  \pm 0.029$     &$\ \ \ 0.065 \pm 0.021 \pm 0.044$ & $     -0.001  \pm 0.013   \pm 0.027$  & $0.06 \pm 0.01$ \\ 
        320--510     & 55  & $   -0.009  \pm 0.020  \pm 0.032$     &$\ \ \ 0.037 \pm 0.025 \pm 0.042$ & $\ \ \ 0.007  \pm 0.015   \pm 0.029$  & $0.05 \pm 0.01$ \\ 
        510--700     & 65  & $   -0.072  \pm 0.045  \pm 0.037$     &$\ \ \ 0.112 \pm 0.059 \pm 0.051$ & $     -0.005  \pm 0.036   \pm 0.034$  & $0.04 \pm 0.01$ \\ 
        700--1000    & 73  & $\ \ \ 0.053  \pm 0.081  \pm 0.039$   &$\ \ \ 0.086 \pm 0.099 \pm 0.068$ & $\ \ \ 0.065  \pm 0.063   \pm 0.040$  & $0.03 \pm 0.01$ \\ 
        $>$1000      & 88  & $   -0.251  \pm 0.140  \pm 0.048$     &$   -0.162 \pm 0.163 \pm 0.105$ & $   -0.219  \pm 0.107   \pm 0.050$      & $0.03 \pm 0.01$ \\ 
        [\cmsTabSkip]
        Inclusive, Mass $>$170  & 44 &  $\ \ \ 0.040 \pm 0.006  \pm 0.015$    & $\ \ \ 0.059 \pm 0.008 \pm 0.019$  & $\ \ \ 0.047 \pm 0.005  \pm 0.013$ & $0.06 \pm 0.01$    \\

    \end{tabular}
  }
\end{table}

\clearpage

\begin{table}[!htb]
  \centering
  \topcaption{Results for the measurement of $\Delta\AFB$ and $\Delta\Aze$ between the muon and electron channels from the maximum likelihood fit to data in different dilepton mass bins
    as well as an inclusive measurement across all mass bins.
    The first and second uncertainties listed with each measurement are statistical and systematic, respectively.}
  \label{tab:delta_afb_results}
  \begin{tabular}{c c c }
    Mass (\GeVns) & $\Delta$\AFB  & $\Delta \Aze$  \\
    \hline

    170--200         &$ -0.045 \pm 0.019 \pm 0.009$          & $\ \ \ 0.018 \pm 0.016   \pm 0.032$    \\ 
    200--250         &$ -0.042 \pm 0.019 \pm 0.006$          & $     -0.027 \pm 0.019  \pm 0.048$   \\ 
    250--320         &$ -0.052 \pm 0.023 \pm 0.006$          & $     -0.092 \pm 0.026  \pm 0.045$   \\ 
    320--510         &$\ \ \ 0.015 \pm 0.023 \pm 0.008$      & $     -0.046 \pm 0.032  \pm 0.045$   \\ 
    510--700         &$ -0.013 \pm 0.043 \pm 0.007$          & $     -0.184 \pm 0.075  \pm 0.053$   \\ 
    700--1000        &$\ \ \ 0.055 \pm 0.064 \pm 0.008$      & $     -0.034 \pm 0.128  \pm 0.068$   \\ 
    $>$ 1000         &$ -0.099 \pm 0.104 \pm 0.014$          & $     -0.090 \pm 0.214  \pm 0.111$   \\ 
    [\cmsTabSkip]
    Inclusive, Mass $>$ 170   & $ -0.026 \pm 0.010  \pm 0.004$    & $     -0.018 \pm 0.011  \pm 0.018$   \\ 

 \end{tabular}
\end{table}

\clearpage

\begin{table}[!htb]
  \centering

  \topcaption{ The fraction of photon-induced background as compared with the total amount of DY signal plus photon-induced events 
    ($N_{\PGg\PGg}/(N_{\PGg\PGg} + N_\mathrm{DY})$) in different dilepton mass bins. 
  These numbers are averaged across the different years and channels.}

  \label{tab:phot_ind_fracs}
  \begin{tabular}{c c c c c c c c }
    Mass (\GeVns) & 170--200 & 200--250 & 250--320 & 320--510 & 510--700 & 700--1000 & $>$ 1000 \\
    \hline
    $\PGg\PGg\to\Pell\Pell$ fraction & 1.8\% & 2.1\% & 2.5\% & 2.8\% & 3.3\% &  3.7\% & 4.1\% \\

  \end{tabular}
\end{table}

In the combined measurement of \AFB, no statistically significant deviations from the SM predictions are observed.
A small difference is found between the muon and electron \AFB's, with $\Delta\AFB$ found to be consistently below zero in the lowest three mass bins,
as well as in the inclusive measurement.
Based on a likelihood scan, the inclusive measurement $\Delta\AFB$ differs from zero at the level of 2.4 standard deviations.

Given that the existence of new gauge bosons would change the asymmetry well below their resonance masses, 
the measurement of \AFB can be used to constrain the existence of heavy \PZpr bosons. 
The constraining power is model dependent, because the interference between the 
\PZpr boson and $\PZ/\PGg^*$ depends on the couplings of the \PZpr boson. 
To give an example of how these measurements can be used to constrain models with \PZpr bosons,
constraints are derived for the sequential standard model (SSM)~\cite{Altarelli:1989ff}. 
The SSM contains a \PZpr boson with the same coupling strength as the SM \PZ boson, up to an overall normalization factor in the left-handed coupling, $\kappa_\mathrm{L}$. 
In order to set 95\% confidence level (\CL) upper limits on properties of the \PZpr boson, a hypothesis test between the SSM and SM is performed based on a comparison of the
predicted values of \AFB in each model and our measurements. 
The test statistic used is the difference in $\chi^2$ between the two models.
Predictions for \AFB in the SSM \PZpr are derived from MC estimates using \MGaMC~\cite{Fuks:2017vtl}. 
Only the measurements of \AFB in the three highest mass bins are used for the $\chi^2$ calculation.
It was checked that the interference from an off-shell several-\TeV SSM \PZpr produces negligible effects in the lower mass bins and their inclusion would not improve the limit.

Figure \ref{fig:Zp_lim} shows the expected and observed 95\% \CL exclusion limits on SSM \PZpr bosons. 
For $\kappa_\mathrm{L} = 1$, which corresponds to a \PZpr boson with exactly the same couplings as the SM \PZ boson, 
a \PZpr with $m_{\PZpr} < 4.4\TeV$ is excluded at 95\% \CL, 
whereas the expected limit was 3.7\TeV. 

Currently these limits are not as strong as those from direct dilepton resonance searches~\cite{CMS:2021ctt, Aad:2019fac}, 
which report limits at 4.90\TeV and 5.15 by ATLAS and CMS, respectively. 
However, the \AFB-based approach is sensitive to wide-width resonances, which may be missed in a search for narrow resonances~\cite{Accomando:2019ahs}. 
Additionally, the sensitivity of the \AFB-based approach will continue to improve with additional data,
with the sensitivity to heavy mass scales expected to scale as the fourth root of the increase in integrated luminosity.
This means limits from the \AFB-based approach should surpass the projected sensitivity of resonance searches~\cite{CidVidal:2018eel}
during the High-Luminosity upgrade of the LHC~\cite{ZurbanoFernandez:2020cco}.

The full tabulated results are provided in HEPData \cite{HEPData}.

\clearpage

\section{Summary}\label{sec:summary}

The CMS detector at the LHC has been used to measure the Drell--Yan forward-backward asymmetry (\AFB) and the angular coefficient \Aze 
as functions of dilepton mass for muon and electron pairs with invariant mass above 170\GeV. 
The measurement is performed using proton-proton collision data collected in 2016--2018 at $\sqrt{s} = 13\TeV$
with an integrated luminosity of 138\fbinv using a template fitting approach.
The combined dimuon and dielectron \AFB measurements show good agreement with the standard model predictions across the full mass range.
An inclusive measurement across the full mass range yields an \AFB of $0.612\pm 0.005\stat\pm 0.007\syst$
and an \Aze of $0.047\pm 0.005\stat\pm 0.013\syst$.
As a test of lepton flavor universality, the difference between the dimuon and dielectron $\AFB\mathrm{s}$ is measured 
and found to agree with zero to within 2.4 standard deviations. 
Using the combined \AFB measurements, limits are set on the existence of additional gauge bosons. 
For a \PZpr boson in the canonical sequential standard model the observed (expected) 95\% confidence level lower limit on the \PZpr mass is 4.4\TeV (3.7 \TeV).

\begin{acknowledgments}
  We congratulate our colleagues in the CERN accelerator departments for the excellent performance of the LHC and thank the technical and administrative staffs at CERN and at other CMS institutes for their contributions to the success of the CMS effort. In addition, we gratefully acknowledge the computing centers and personnel of the Worldwide LHC Computing Grid and other centers for delivering so effectively the computing infrastructure essential to our analyses. Finally, we acknowledge the enduring support for the construction and operation of the LHC, the CMS detector, and the supporting computing infrastructure provided by the following funding agencies: BMBWF and FWF (Austria); FNRS and FWO (Belgium); CNPq, CAPES, FAPERJ, FAPERGS, and FAPESP (Brazil); MES (Bulgaria); CERN; CAS, MoST, and NSFC (China); MINCIENCIAS (Colombia); MSES and CSF (Croatia); RIF (Cyprus); SENESCYT (Ecuador); MoER, ERC PUT and ERDF (Estonia); Academy of Finland, MEC, and HIP (Finland); CEA and CNRS/IN2P3 (France); BMBF, DFG, and HGF (Germany); GSRT (Greece); NKFIA (Hungary); DAE and DST (India); IPM (Iran); SFI (Ireland); INFN (Italy); MSIP and NRF (Republic of Korea); MES (Latvia); LAS (Lithuania); MOE and UM (Malaysia); BUAP, CINVESTAV, CONACYT, LNS, SEP, and UASLP-FAI (Mexico); MOS (Montenegro); MBIE (New Zealand); PAEC (Pakistan); MSHE and NSC (Poland); FCT (Portugal); JINR (Dubna); MON, RosAtom, RAS, RFBR, and NRC KI (Russia); MESTD (Serbia); SEIDI, CPAN, PCTI, and FEDER (Spain); MOSTR (Sri Lanka); Swiss Funding Agencies (Switzerland); MST (Taipei); ThEPCenter, IPST, STAR, and NSTDA (Thailand); TUBITAK and TAEK (Turkey); NASU (Ukraine); STFC (United Kingdom); DOE and NSF (USA).

\hyphenation{Rachada-pisek} Individuals have received support from the Marie-Curie program and the European Research Council and Horizon 2020 Grant, contract Nos.\ 675440, 724704, 752730, 765710 and 824093 (European Union); the Leventis Foundation; the Alfred P.\ Sloan Foundation; the Alexander von Humboldt Foundation; the Belgian Federal Science Policy Office; the Fonds pour la Formation \`a la Recherche dans l'Industrie et dans l'Agriculture (FRIA-Belgium); the Agentschap voor Innovatie door Wetenschap en Technologie (IWT-Belgium); the F.R.S.-FNRS and FWO (Belgium) under the ``Excellence of Science -- EOS" -- be.h project n.\ 30820817; the Beijing Municipal Science \& Technology Commission, No. Z191100007219010; the Ministry of Education, Youth and Sports (MEYS) of the Czech Republic; the Deutsche Forschungsgemeinschaft (DFG), under Germany's Excellence Strategy -- EXC 2121 ``Quantum Universe" -- 390833306, and under project number 400140256 - GRK2497; the Lend\"ulet (``Momentum") Program and the J\'anos Bolyai Research Scholarship of the Hungarian Academy of Sciences, the New National Excellence Program \'UNKP, the NKFIA research grants 123842, 123959, 124845, 124850, 125105, 128713, 128786, and 129058 (Hungary); the Council of Science and Industrial Research, India; the Latvian Council of Science; the Ministry of Science and Higher Education and the National Science Center, contracts Opus 2014/15/B/ST2/03998 and 2015/19/B/ST2/02861 (Poland); the National Priorities Research Program by Qatar National Research Fund; the Ministry of Science and Higher Education, project no. 0723-2020-0041 (Russia); the Programa Estatal de Fomento de la Investigaci{\'o}n Cient{\'i}fica y T{\'e}cnica de Excelencia Mar\'{\i}a de Maeztu, grant MDM-2015-0509 and the Programa Severo Ochoa del Principado de Asturias; the Thalis and Aristeia programs cofinanced by EU-ESF and the Greek NSRF; the Rachadapisek Sompot Fund for Postdoctoral Fellowship, Chulalongkorn University and the Chulalongkorn Academic into Its 2nd Century Project Advancement Project (Thailand); the Kavli Foundation; the Nvidia Corporation; the SuperMicro Corporation; the Welch Foundation, contract C-1845; and the Weston Havens Foundation (USA).
\end{acknowledgments}

\bibliography{auto_generated} 

\cleardoublepage \appendix\section{The CMS Collaboration \label{app:collab}}\begin{sloppypar}\hyphenpenalty=5000\widowpenalty=500\clubpenalty=5000\cmsinstitute{Yerevan~Physics~Institute, Yerevan, Armenia}
A.~Tumasyan
\cmsinstitute{Institut~f\"{u}r~Hochenergiephysik, Vienna, Austria}
W.~Adam\cmsorcid{0000-0001-9099-4341}, J.W.~Andrejkovic, T.~Bergauer\cmsorcid{0000-0002-5786-0293}, S.~Chatterjee\cmsorcid{0000-0003-2660-0349}, K.~Damanakis, M.~Dragicevic\cmsorcid{0000-0003-1967-6783}, A.~Escalante~Del~Valle\cmsorcid{0000-0002-9702-6359}, R.~Fr\"{u}hwirth\cmsAuthorMark{1}, M.~Jeitler\cmsAuthorMark{1}\cmsorcid{0000-0002-5141-9560}, N.~Krammer, L.~Lechner\cmsorcid{0000-0002-3065-1141}, D.~Liko, I.~Mikulec, P.~Paulitsch, F.M.~Pitters, J.~Schieck\cmsAuthorMark{1}\cmsorcid{0000-0002-1058-8093}, R.~Sch\"{o}fbeck\cmsorcid{0000-0002-2332-8784}, D.~Schwarz, S.~Templ\cmsorcid{0000-0003-3137-5692}, W.~Waltenberger\cmsorcid{0000-0002-6215-7228}, C.-E.~Wulz\cmsAuthorMark{1}\cmsorcid{0000-0001-9226-5812}
% ...
\vskip\cmsinstskip
% ...
1:~Also at TU Wien, Wien, Austria\\
% ...
\end{sloppypar}
\end{document}